\documentclass[11pt,preprint]{aastex}

\newcommand{\etal}{{\it et~al.}}

\usepackage{longtable}

\begin{document}

\title{Physical Properties of 299 NEOs Manually Recovered in Over Five Years of NEOWISE Survey Data}

\author{Joseph R. Masiero\altaffilmark{1}, Patrice Smith\altaffilmark{1,2}, Lean D. Teodoro\altaffilmark{1,3}, A.K. Mainzer\altaffilmark{4}, R.M. Cutri\altaffilmark{5}, T. Grav\altaffilmark{4}, E. L. Wright\altaffilmark{6}}

\altaffiltext{1}{Jet Propulsion Laboratory/California Institute of Technology, 4800 Oak Grove Dr., MS 183-301, Pasadena, CA 91109, USA, {\it Joseph.Masiero@jpl.nasa.gov}}
\altaffiltext{2}{University of Hawaii, Hilo, HI 96720 USA}
\altaffiltext{3}{University of Hawaii, Manoa, HI 96822 USA}
\altaffiltext{4}{University of Arizona, Tucson, AZ 85721 USA}
\altaffiltext{5}{California Institute of Technology, IPAC, 1200 California Blvd, Pasadena, CA 91125 USA}
\altaffiltext{6}{University of California, Los Angeles, CA, 90095}

\begin{abstract}

  Thermal infrared measurements of near-Earth objects provide critical
  data for constraining their physical properties such as size.  The
  NEOWISE mission has been conducting an all-sky infrared survey to
  gather such data and improve our understanding of this population.
  While automated routines are employed to identify the majority of
  moving objects detected by NEOWISE, a subset of objects will have
  dynamical properties that fall outside the window detectable to
  these routines.  Using the population of known near-Earth objects,
  we have conducted a manual search for detections of these objects
  that were previously unreported.  We report $303$ new epochs of
  observations for $299$ unique near-Earth objects of which $239$ have
  no previous physical property characterization from the NEOWISE
  Reactivation mission.  As these objects are drawn from a list with
  inherent optical selection biases, the distribution of measured
  albedos is skewed to higher values than is seen for the
  diameter-selected population detected by the automated routines.
  These results demonstrate the importance and benefit of periodic
  searches of the archival NEOWISE data.

\end{abstract}

\section{Introduction}

Small bodies of the Solar System with perihelia less than $1.3~$AU are
known as near-Earth objects (NEOs).  These objects are warmed by
incident sunlight, and re-emit that light as thermal infrared
emission, with objects that are closer to the Sun being warmer and
thus brighter at infrared wavelengths.  The Near-Earth Object
Wide-field Infrared Survey Explorer (NEOWISE) has been carrying out a
survey of the sky at thermal infrared wavelengths to detect and
characterize these NEOs \citep{mainzer14neowise}.  NEOWISE began its
survey on 13 December 2013 after the reactivation of the Wide-field
Infrared Survey Explorer spacecraft \citep{wright10,mainzer11}, and
has continued surveying for over 6 years at $3.4~\mu$m and
  $4.6~\mu$m (referred to as W1 and W2, respectively).  The images
obtained by NEOWISE are automatically scanned for detections of moving
Solar System objects, and these detections are reported regularly to
the Minor Planet Center as part of regular survey data processing.
Thermal modeling can be performed on these detections allowing for
diameters to be constrained, as well as albedos when visible light
measurements are also available, and these parameters for objects
observed during the NEOWISE reactivation mission have been described
in a series of papers
\citep{mainzer14neowise,nugent15,nugent16,masiero17,masieroY45}.

The automated WISE Moving Object Processing System (WMOPS) searches
the NEOWISE data within a set of bounds that allow it to detect most,
but not all, NEOs passing through the field of view.  Tracklets are
built from chains of detections, within set limits on acceleration,
changing direction of motion, and minimum number of observations.
These limits are set to maximize the number of objects identified
while maintaining a reasonable number of false-positives sent for
human quality assurance \citep[cf.][]{cutri15}.  These limits will,
however, mean that some objects of interest will not be identified.
Specifically, NEOs passing close to NEOWISE, and thus having a high
rate of motion through the field of view, are less likely to meet the
threshold of the minimum of $5$ detections to be identified
automatically.  Other objects, based on viewing geometry, will exceed
the allowable changes in rate and direction of motion for linking.

Because NEOWISE archives all full-frame images acquired during the
survey and a database of sources detected in those images, it is
possible to conduct a search for known NEOs missed by the automated
processing after the fact.  We present in this work a search for these
objects, using the list of all currently known NEOs as of 1 June 2019
as input.  A previous search to this effect was performed by
\citet{masiero18}, covering the first three years of the NEOWISE
Reactivated survey, while \citet{mainzer14tinyneo} presented a similar
search of the data from the cryogenic portion of the original WISE
survey.  This work uses the larger list of NEOs known presently, as
well as all data from the first five and a half years of the NEOWISE
Reactivation data.  The aim of this search is to increase the number
of NEOs with diameter and albedo characterization in order to expand
our knowledge of this population, and make best use of the data
obtained by the NEOWISE mission.

\section{Methods}

We used for our search the list of all known NEOs recorded in the
Minor Planet Center's (MPC) orbital element list MPCORB\footnote{\it
  https://www.minorplanetcenter.net} as of June 1, 2019.  Using these
orbital elements, we determined the subset of objects that were at
Solar elongations between $88^\circ - 115^\circ$ (the NEOWISE field of
regard during the mission) and predicted apparent magnitudes of
$V<20~$mag during the time of the NEOWISE Reactivation survey.  Each
object on this list was queried from one month before to one month
after the date of peak brightness in the searched elongation region
using the IRSA WISE Moving Object Search Tool\footnote{\it
  https://irsa.ipac.caltech.edu/applications/MOST/} through the API
interface.  These searches produced a list of WISE images that
overlapped the predicted position of each NEO, and thus might contain
a previously unidentified detection.

The images were visually inspected to look for sources coincident with
the predicted positions of each object.  Images without identifiable
sources were removed from the list, and then the NEOWISE-R Single
Exposure Source Table\footnote{\it https://irsa.ipac.caltech.edu/} was
searched for entries within $5''$ of the predicted position of each
NEO at the time of the NEOWISE observation.  Objects with NEOWISE
observations already reported to the MPC by the automated WMOPS system
were not included in this search for the epochs that had been
reported.  However, this search did return additional detections for
previously reported NEOs at new epochs.  Increasing the number of
identified observing epochs for NEOs is critical for advanced
thermophysical modeling work that uses multiple viewing geometries to
constrain NEO surface thermal inertias
\citep[e.g.][etc.]{delbo07,koren15,delbo15,hanus16,masiero19}.

The resulting list of returned detections contained objects with as
few as one detection to as many as 34 detections.  For objects
detected in only a single image, there is a significant potential for
stars, cosmic rays, or other artifacts and noise sources to masquerade
as real detections, so a second visual inspection of these detections
was carried out.  After this inspection, $48$ NEOs with single
detections were determined to have a high probability of being real,
and $38$ of these were detected in both NEOWISE bandpasses (which are
imaged simultaneously).

\begin{figure}[ht]
\begin{center}
\includegraphics[scale=0.6]{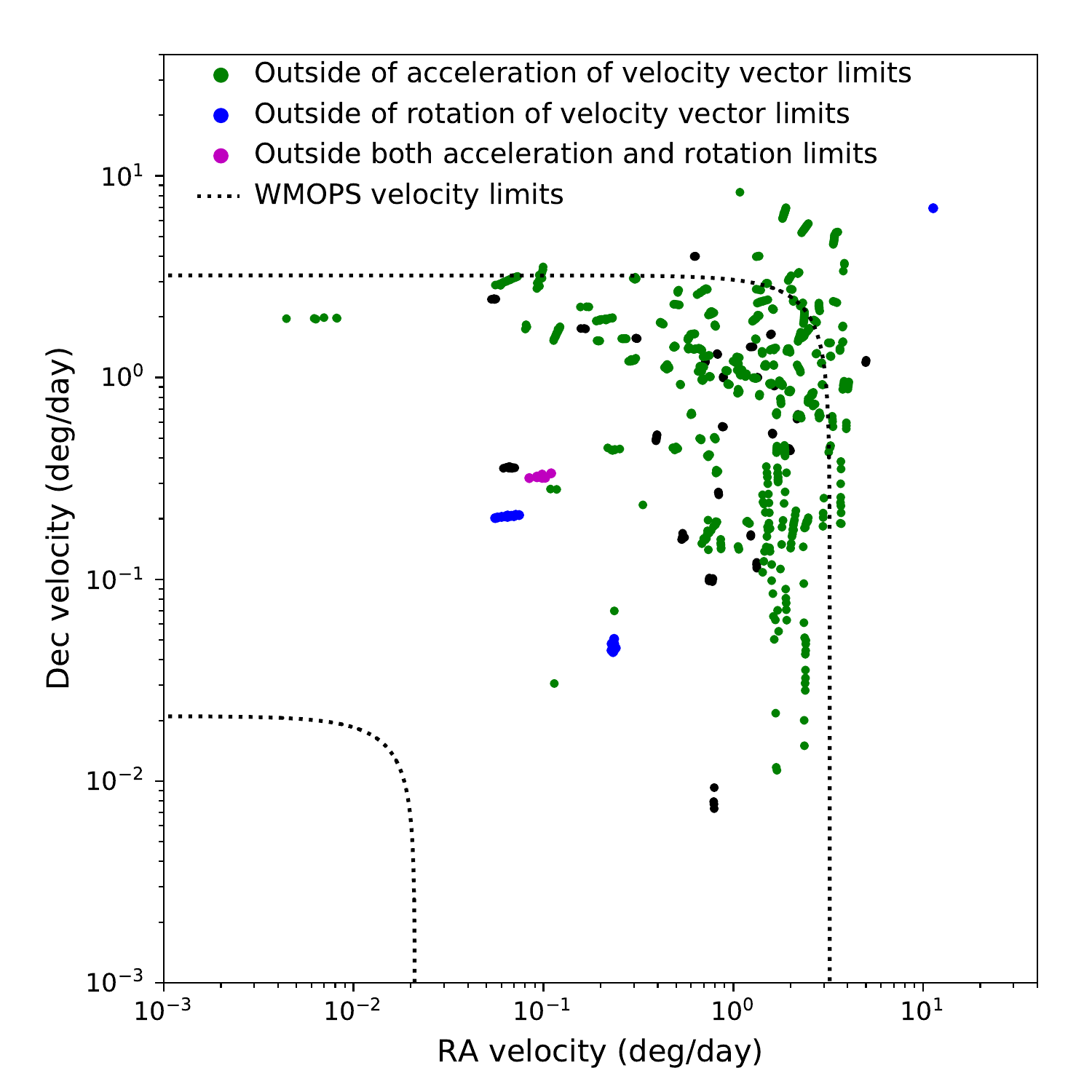} \protect\caption{
  Rates of motion in RA and Dec of all objects presented here that
  were detected more than five times (the minimum required for
  automated detection).  Each tracklet will be shown as a string of
  dots based on the changes in motion within the tracklet.  Dotted
  lines show the lower and upper limits to rate of motion that WMOPS
  will detect.  Green points are tracklets that had accelerations beyond
  the limit accepted by WMOPS; blue point had rotations in the angle
  of motion outside the WMOPS limit; purple points are tracklets the
  exceeded both limits.  Black points within the WMOPS velocity limits
  were not found either due to falling below the detection threshold,
  or having detections rejected by the atlas-based stationary object
  rejection routines.  }
\label{fig.rom}
\end{center}
\end{figure}

Objects detected five or more times are of particular interest because
they represent tracklets that could have been identified by the WMOPS
automated processing routines, but were not for various reasons.  The
WMOPS software identifies objects by first creating pairs of
detections, and then linking these pairs based on common motion
vectors.  The WMOPS velocity limits on creating pairs ($0.021~$deg
day$^{-1} <$velocity$<3.22~$deg day$^{-1}$) along with the
acceleration tolerance ($<0.01~$deg day$^{-2}$) and angle of motion
tolerance ($<1^\circ$) used for linking them define a phase space
where WMOPS can detect objects.  The majority of the NEOs with five or
more detections found in our manual search here are outside this phase
space, as shown in Figure~\ref{fig.rom}.  The remaining $21$ objects
that are within this phase space were lost either because they were
near the signal-to-noise detection limit (S/N$>4.5$) used to compile
the input list to build detection pairs, or they were in a region of
sky with a dense background and so were lost during the catalog-based
filtering for stationary object rejection that is done at the
beginning of WMOPS processing.

It is important to note that the vast majority of NEOs were discovered
by ground-based visible light telescopes, and thus will have a
preferential bias toward high-albedo objects at any given size range,
as visible surveys are brightness-limited.  This bias will combine
with the selection bias we impart by looking for objects with visually
bright apparitions at the time they approached the NEOWISE field of
regard, resulting in our list preferentially containing more
high-albedo asteroids than would be found in an unbiased sample of the
true NEO population.

\section{Thermal Modeling}

We employed the Near-Earth Asteroid Thermal Model
\citep[NEATM,][]{harris98} to constrain the physical properties of the
detected asteroids using their NEOWISE thermal infrared measurements,
following the procedure used in our previous study of NEOWISE
manually-recovered NEOs \citep{masiero18}.  Using the tables of
detections published in IRSA, we extract the W1 and W2 profile-fit
photometry and astrometry for the detections of these objects.  In
addition to visual inspection of all detections to remove
contamination from cosmic rays, artifacts, and background stars, we
also performed an automated filtering on the detections before using
them for thermal modeling, as described below.

Every detection published in IRSA includes a reduced $\chi^2$
($rchi2$) value of the fit of the PSF model to the presumed source in
each bandpass.  We reject all detections with $rchi2>20$, as this
indicates a high likelihood of a contaminated detection.  In general,
values of $rchi2>5$ are removed from analysis as they have a higher
likelihood of being cosmic rays \citep[e.g.][]{masieroY45}.  However
any source that is trailed, even at the sub-PSF level, may have an
increased $rchi2$ value and so for this work we only reject sources
with $rchi2>20$ to remove the most serious cosmic ray contamination.
This cut was determined through visual validation of the detections of
singleton objects that were slightly trailed and seen in both bands,
providing a guide for what the largest acceptable $rchi2$ would be for
this work.

We also reject 2-band detections where $W1-W2<1~$mag to remove
contamination from stars.  During initial thermal modeling tests, we
found instances of detections contaminated by comparably bright stars,
which created problems for the thermal model fitting.  The majority of
stars are in the Rayleigh-Jeans portion of their spectral energy
distribution at the W1 and W2 wavelengths, so the expected W1-W2 color
is $\sim0$, while for sources with rising thermal emission in W2, such
as asteroids inside $\sim3~$AU, this color is expected to be redder.
An analysis of the W1-W2 colors of all sources detected in two bands
shows a peak of the distribution at $W1-W2\sim2.6~$mag, with the vast
majority of sources within $1.5~$mag of that peak.  Visual inspection
of sources with $W1-W2<1~$mag confirmed the predicted positions
coincide with stationary background objects, so this color is used as
a cut on the data prior to thermal modeling.  In total, $40$
detections (out of $1688$) were eliminated from fitting by color and
$rchi2$ cuts. This cut would also reject distant asteroids dominated
by reflected light in the W1 and W2 bands, however these are highly
unlikely to have large rates of motion and be missed by WMOPS, if
bright enough to be detected.

Proximity of the asteroid to the WISE spacecraft is the main reason
that objects will be moving too fast or accelerating too much to be
detected often enough for automated identification.  As NEOWISE
observes at a narrow range of Solar elongations, observer distance and
observational phase angle are coupled.  This results in the detections
presented here being made at higher phase angles than NEOs discussed
in previous work \citep{nugent15,nugent16,masiero17}.  The NEATM
beaming parameter, used to account for model uncertainties, is
correlated with phase \citep{mainzer11neo}, and as such we assume a
larger beaming parameter here than is used for other studies of
NEOWISE-observed NEOs.  Following \citet{masiero18} we assume model
beaming parameters of $\eta=2.0\pm0.5$.  We also assume ratios of the
infrared albedo at $3.4~\mu$m ($p_{IR}$) to the visual albedo of
$p_{IR}/p_V = 1.6\pm1.0$.  The uncertainties on these parameters are
fed into our Monte Carlo analysis to determine overall diameter
uncertainty.

We employ a Monte Carlo analysis of our fit to constrain the
statistical uncertainty on our diameter determinations. Taking the
uncertainties on our measured parameters as well as the assumed
uncertainty on our fixed model parameters as discussed above, we vary
each input to our model over 25 iterations and use the resultant
spread of the diameters and albedos in the model solutions as the
quoted uncertainty on our best-fit values.  For the $W1$ and $W2$
magnitudes, measurement uncertainties are taken from the
Single-Exposure Source Table.  For the measured $H$ magnitude, we
assume an uncertainty of $0.2~$ magnitudes following previous work
\citep[e.g.][]{masiero18}, as no uncertainty is provided in the MPC
catalog.  We assume all parameters can be modeled by a Gaussian
distribution, with the uncertainties giving the $1\sigma$ value and
the measured/assumed value as the mean of the distribution.  We note
that for the flux uncertainties specifically, \citet{wright18} showed
that while they are not strictly Gaussian, the actual measurement
uncertainty can be encompassed by a Gaussian using the published
value.  Thus, this assumption is sufficient for our analysis where the
fit uncertainties are dominated by other terms, such as the unknown
beaming parameter.

In addition to statistical uncertainties on the thermal model fits,
there are also systematic model uncertainties to consider.  NEATM is
an imperfect model that assumes the night side of an asteroid
contributes no thermal emission.  While this will only have a small
effect on objects at low phase angles, at higher phases this results
in an underestimation of the emitted flux and an overestimation of the
diameter.  \citet{mommert18} showed that beyond a phase angle of
$\alpha>65^\circ$ NEATM deviates from the true diameter, and requires
a correction factor to accurately reproduce input thermophysical model
parameters.  The objects in our sample have a mean phase angle of
$\alpha=73^\circ$, and three-quarters of them were seen at phases
beyond $\alpha=65^\circ$.  In light of this, we also calculated the
corrected diameters and albedos based on the correction equations from
\citet{mommert18}.  However we note that this correction was developed
from a NEATM model fit to multiple wavelengths spanning the peak of
thermal emission, and so may not completely correct for the model
offsets in our implementation of NEATM.

A further source of uncertainty for the fits presented in this work is
the unknown light curve phase for objects with a small number of
detections.  As discussed in \citet{mainzer14tinyneo} and
\citet{masiero18}, when an object has only a few samplings of its
light curve available, the error on determining the mean of the light
curve (which enables fitting an effective spherical diameter for a
body) increases, up to $\sim30\%$ of the light curve amplitude for a
single detection.  For an object with a light curve amplitude of
$>1~$mag this can result in an offset of the fitted diameter from the
true spherical equivalent diameter of $>15\%$.  In addition to this
effect, light curve variations as well as the Eddington bias can
result in objects near the detection limit having overestimated
fluxes, artificially increasing the fitted diameter compared to the
true size.  For the objects presented here, our median S/N in $W2$ was
larger than 10, so this will not have a significant impact on these
fits, but should be kept in mind when dealing with objects with a
small number of detections.

\section{Results and Discussion}

The results of our modeling, including the best fit values and Monte
Carlo error analyses, are presented in Table~\ref{tab.props}.  We
describe $303$ NEATM fits of $299$ unique near-Earth objects, of which
$239$ had not been previously characterized by the reactived NEOWISE
mission.  The remaining $60$ objects had been detected and reported by
the spacecraft at a different observing epoch since the reactivation
in Dec 2013.  The four objects with multiple fits show good agreement
within the quoted statistical uncertainties from the Monte Carlo
analysis.  For all objects and epochs, the measured astrometry data
have been reported to the MPC, and are archived there.

Previous work has shown that fits of objects with reflected light
contributions to the $W2$ band above $10\%$ of the total flux are
less-reliable \citep[cf.][]{masiero17}, and thus we removed from our
physical property list fits of $19$ objects that were detected by
NEOWISE but fell in this regime.  These objects were:
(152952),(163132),(163243),(281375),(304640),(364136),(388945),
(418198), (472263), (530743), (536531), 2009 FU23, 2017 OO1, 2017
RV17, 2017 VX1, 2017 VW13, 2018 RP8, 2018 UY, 2019 CE.  The
astrometry for these objects recovered in our search were
still submitted to the MPC.

\begin{figure}[ht]
 \begin{center}
 \includegraphics[scale=0.5]{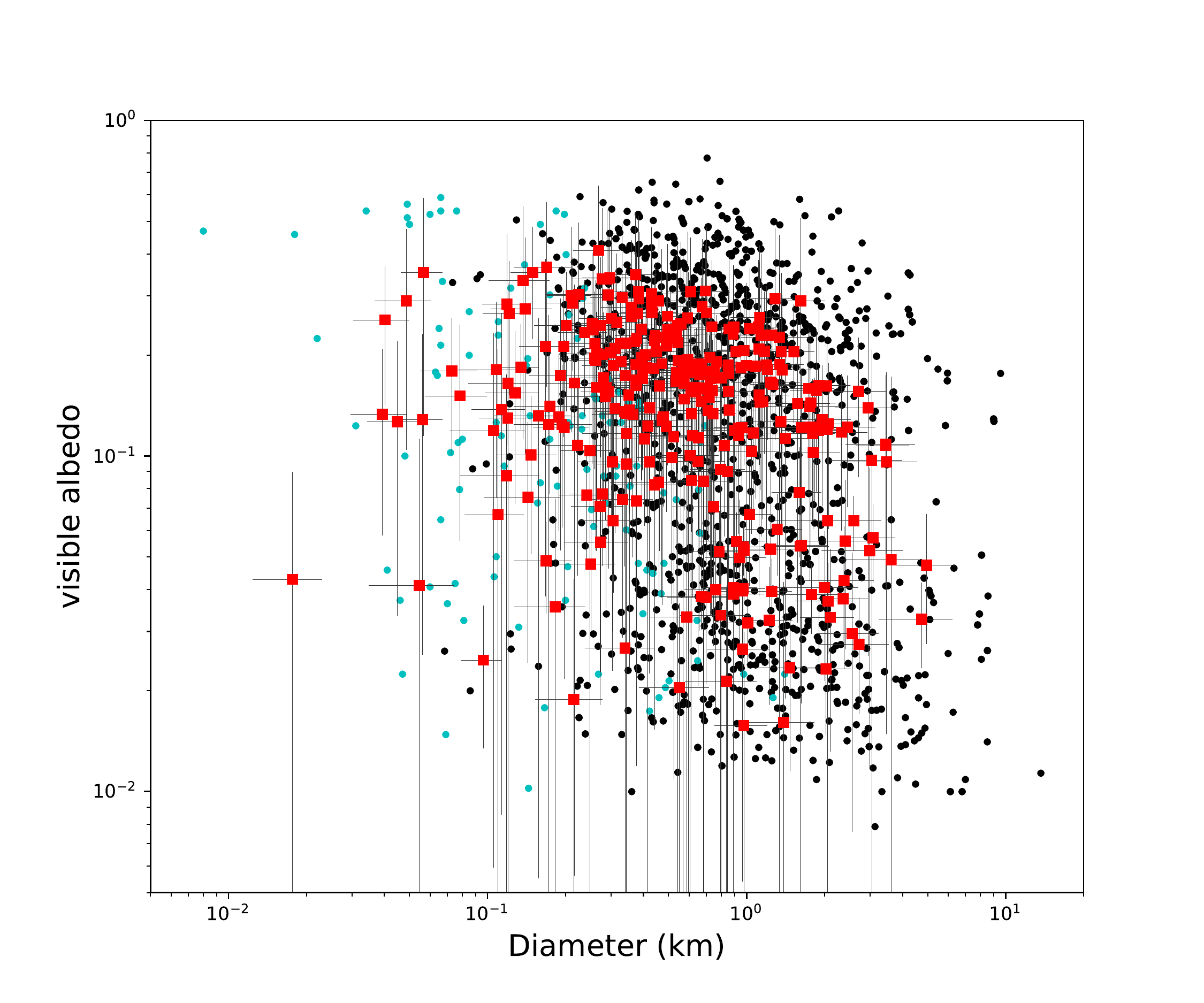}
 \protect\caption{Diameters and albedos for near-Earth objects
   presented here (red squares), compared with manually detected NEOs
   presented in \citet{masiero18} (cyan dots) and NEOs detected
   automatically in the first five years of the NEOWISE Reactivation
   survey (black dots).  Error bars are shown only for the newly
   presented objects, but are comparable in size for all fits.  The
   search criteria used here that targets objects with predicted
   visual magnitudes brighter than V$=20~$mag at the time they passed
   through the NEOWISE field of regard, combined with the preferential
   discovery of high albedo objects by optical telescopes, results in a
   significant bias against low albedo NEOs. }
 \label{fig.diamalb}
 \end{center}
 \end{figure}

We show in Figure~\ref{fig.diamalb} a comparison of the diameters and
albedos for the objects presented here, along with those published
from our previous manual recovery search \citep{masiero18} and the
fits for objects found by our automated WMOPS detection algorithms.
As expected, the bias in favor of high albedo objects (due to the
initial discovery selection effect combined with the selection effect
on the input list for our search) is clearly apparent in the
distribution of our sample, with the majority of objects having fits
in the with albedos greater than $p_V>0.1$.  This work recovered more
large objects than our previous manual search, with $78$ objects
having fitted diameters larger than $1~$km.  This is because our new
search included all objects in the NEO orbital list, while our
previous work focused on short-arc asteroids that had been discovered
more recently.  More recently discovered objects tend to be smaller as
there are fewer large objects remaining undiscovered as the global NEO
survey programs progress.

\begin{figure}[ht]
 \begin{center}
 \includegraphics[scale=0.5]{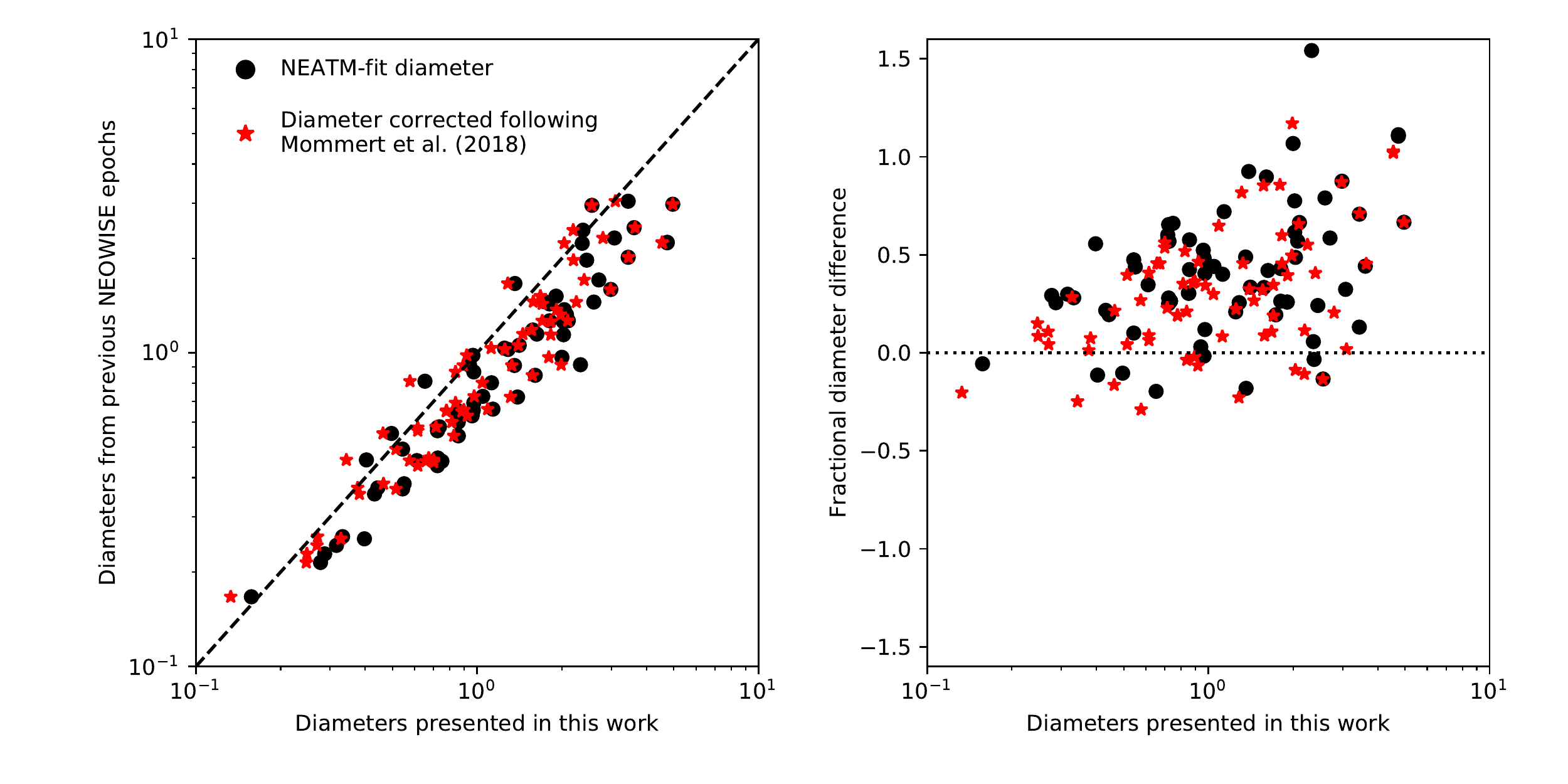}
 \protect\caption{Comparison of the diameters presented in this work
   to previously-published NEOWISE diameters of the same NEOs at
   different observing epochs (left).  Fits from the NEATM model are
   shown as black circles, and sizes corrected following
   \citep{mommert18} are shown as red stars.  The fractional
   difference between these fits (right) shows that the NEATM fits to
   the higher phase angle data tends to over-estimate the sizes
   compared to previous work. }
 \label{fig.comp}
 \end{center}
 \end{figure}

We can use the $60$ objects that have previously reported
NEOWISE-Reactivation diameters to verify the accuracy of the model
results presented here.  We show in Figure~\ref{fig.comp} the
comparison of the diameters presented in this paper to those
previously published values.  The diameters in this work show a fairly
large random scatter, as well as a systematic offset to larger values.
This is likely due to a combination of effects including the smaller
number of detections which present a bias toward light curve maxima,
as well as the larger phase angles of observation which are
detrimental to the NEATM fitting accuracy.  Although we include the
correction to the NEATM fits proposed by \citet{mommert18}, which
provides a small improvement to this offset, it does not fully
eliminate it.  

\begin{figure}[ht]
 \begin{center}
 \includegraphics[scale=0.5]{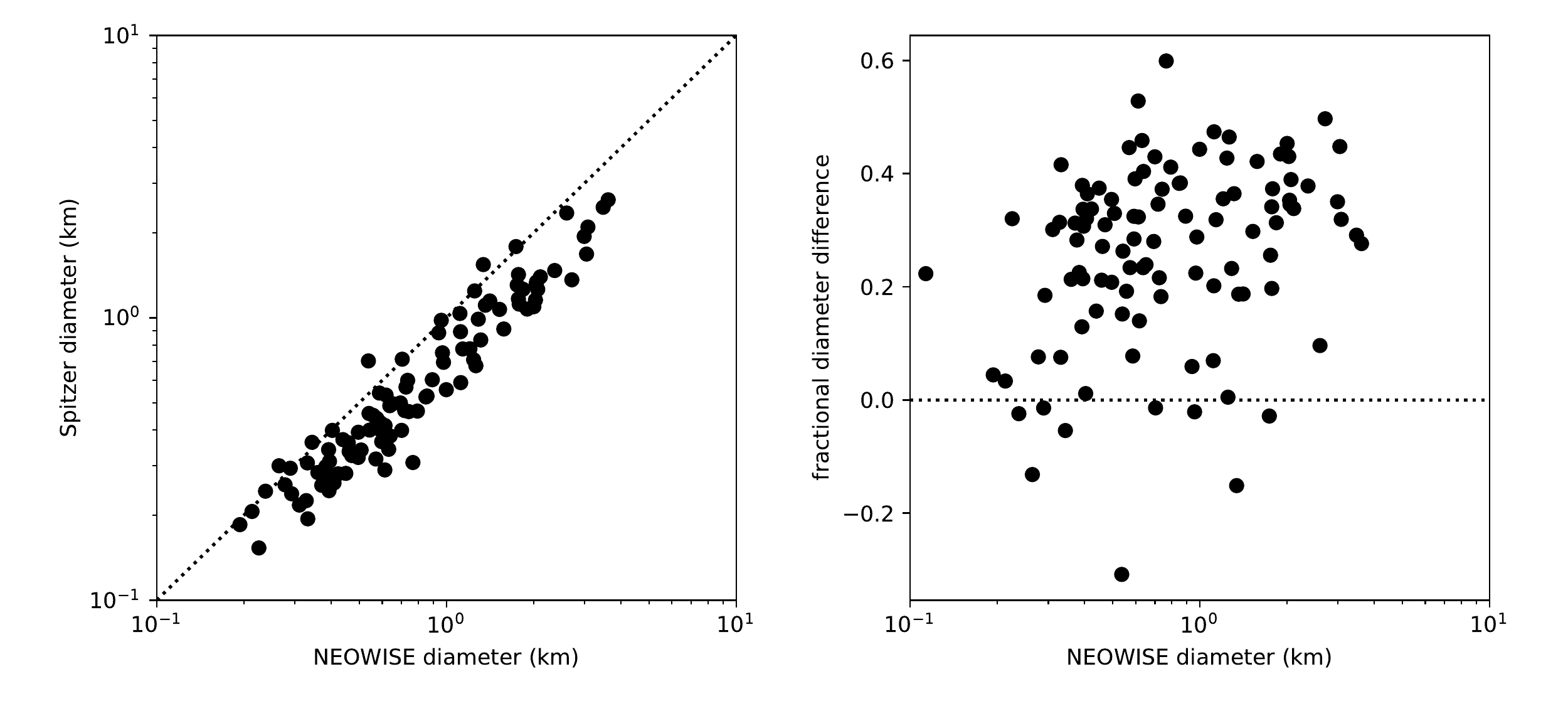}
 \protect\caption{Comparison of NEO diameters presented here to sizes
   determined by Spitzer (left) and fractional diameter difference
   between the two data sets (right).  Sizes derived here are larger
   than those found by Spitzer by $\sim30\%$, which is a result of the
   larger assumed beaming parameter chosen here. }
 \label{fig.spitzer}
 \end{center}
 \end{figure}

We can also compare our physical property results to those NEOs with
sizes measured by the Spitzer space telescope \citep{trilling16}.  The
objects targeted by Spitzer are generally are smaller than those
regularly detected by the NEOWISE automated pipeline, but our manual
recovery allows us to find more objects in this smaller size range.
We extracted all NEOs with Spitzer CH2 S/N $> 5$ from the online
database of fitted
properties\footnote{\it{http://nearearthobjects.nau.edu}}. There are
105 objects from there in common with our table of fitted properties.
A comparison between the sizes published in these two data sets is
shown in Figure~\ref{fig.spitzer}.

The fits presented here here tend to be somewhat larger than those
derived from the Spitzer data, though not as large as the offset from
previous NEOwISE epochs.  Due to the limited time window over which
Spitzer observed most NEOs, those data may show some of the same
effects and biases associated with limited lightcurve sampling as our
fits, however this effect will be more pronounced for objects seen by
NEOWISE in a small number of short exposures.  Additionally, the
larger beaming parameter used for our fits (due to the typically
higher phase angle of observation) also will skew the fits to a larger
size.

Given the results of these comparisons, the diameters presented here
may be over-estimates of the actual spherical equivalent diameter.
This highlights the limitations of NEATM as well as the need to use
thermophysical models for objects with high phase angle observations
to more accurately constrain sizes.  Performing thermophysical
modeling on the NEOWISE and Spitzer observations together would
further enhance the benefits of this technique.

\section{Conclusions}

We present diameter fits for $299$ NEOs found through manual searches
of the NEOWISE data archive, $239$ of which had no previously reported
NEOWISE-derived diameter.  Due to the detection circumstances of these
objects, and the larger phase angles at which they are observed
compared to the automatically-detected objects, the uncertainty on the
diameter determination is larger, and there is a systematic offset in
the fitted diameters to larger sizes in this work.  Thus, the
diameters presented in many cases represent over-estimates of the true
effective spherical diameter of these bodies.  These observations,
however, do provide some constraint on the overall size of these NEOs,
and are important additions to the multiple observing epochs needed to
carry out more advanced thermophysical modeling.

When combined with all previous publications, the diameters presented
here bring the total number of NEOs characterized by the
NEOWISE-Reactivation mission to $1193$ since the survey was restarted
in Dec 2013.  Combined with objects observed during the initial WISE
mission phases before hibernation, a total of $1652$ NEOs have
physical property characterization using data from the WISE and
NEOWISE missions.  As NEOs continue to be discovered by ongoing and
future surveys, the archived NEOWISE single-frame images will be an
important resource for recovering characterization data well after the
mission has ended.  The NEOWISE data are already an important legacy
data set with a large amount of information not currently known
that is waiting to be revealed.

\section*{Acknowledgments}
The research was carried out at the Jet Propulsion Laboratory,
California Institute of Technology, under a contract with the National
Aeronautics and Space Administration (80NM0018D004).  This publication
makes use of data products from the Wide-field Infrared Survey
Explorer, which is a joint project of the University of California,
Los Angeles, and the Jet Propulsion Laboratory/California Institute of
Technology, funded by the National Aeronautics and Space
Administration.  This publication also makes use of data products from
NEOWISE, which is a joint project of the University of Arizona and Jet
Propulsion Laboratory/California Institute of Technology, funded by
the Planetary Science Division of the National Aeronautics and Space
Administration.  This research has made use of data and services
provided by the International Astronomical Union's Minor Planet
Center.  This research has made use of the NASA/IPAC Infrared Science
Archive, which is operated by the California Institute of Technology,
under contract with the National Aeronautics and Space Administration.
This research has made extensive use of the {\it numpy}, {\it scipy}
\citep{scipy}, {\it astropy} \citep{astropy1, astropy2}, and {\it
  matplotlib} Python packages.

\vspace{0.5in}

\scriptsize{
\begin{longtable}{lccccccccccc}
\caption{Thermal model fits for manually recovered NEOs detected in
  the NEOWISE Reactivation survey data.  Names are in MPC-packed
  format, H and G are the input photometric parameter measurements
  used by the model, $p_V$ is the visible light albedo, and n$_{W1}$
  and n$_{W2}$ are the numbers of detections in the W1 and W2
  bandpasses. $D_{corr}$ and ${p_V}_{corr}$ have been corrected
  following the equations in \citet{mommert18}. \\$^{\dagger}$Albedo
  uncertainties are symmetric in log-space as the error is dominated
  by the uncertainty on $H$; the asymmetric linear equivalents of the
  $1 \sigma$ log-space uncertainties are presented here.}\\

  Name  &  input H  &   G   &   Diameter  &  D$_{corr}$ &  ${p_V}^\dagger$  &  ${p_V}_{corr}$ & beaming  &  n$_{W1}$  &n$_{W2}$ & phase  & Mean MJD \\
         & (mag)     &       &    (km)     &      (km)  &                  &                &          &           &         &  (deg) & days \\
   \hline
   \endhead
  85275 & 16.10 &  0.15 & 3.614 $\pm$ 1.475 & 3.643 & 0.049 (+0.123/-0.035) & 0.048 & 2.00 $\pm$ 0.50 &   0 &   1 & 37.11 & 57329.3314\\
  85713 & 15.60 &  0.15 & 3.442 $\pm$ 1.028 & 3.098 & 0.108 (+0.074/-0.044) & 0.134 & 2.00 $\pm$ 0.50 &   5 &   5 & 75.50 & 56977.7030\\
  85953 & 18.10 &  0.15 & 0.959 $\pm$ 0.207 & 0.922 & 0.122 (+0.058/-0.039) & 0.134 & 2.00 $\pm$ 0.50 &   3 &   3 & 63.29 & 58220.0983\\
  85989 & 17.10 &  0.15 & 2.986 $\pm$ 1.041 & 2.978 & 0.052 (+0.043/-0.023) & 0.053 & 2.00 $\pm$ 0.50 &  13 &  13 & 49.39 & 56875.8718\\
  85990 & 20.20 &  0.15 & 0.404 $\pm$ 0.108 & 0.342 & 0.113 (+0.069/-0.043) & 0.153 & 2.00 $\pm$ 0.50 &   2 &   2 & 83.11 & 57023.2583\\
  86819 & 17.40 &  0.15 & 1.124 $\pm$ 0.259 & 1.043 & 0.259 (+0.165/-0.101) & 0.304 & 2.00 $\pm$ 0.50 &  16 &  17 & 70.67 & 56934.4738\\
  88213 & 19.40 &  0.15 & 0.940 $\pm$ 0.249 & 0.893 & 0.050 (+0.030/-0.019) & 0.056 & 2.00 $\pm$ 0.50 &   6 &   6 & 66.03 & 58195.2370\\
  89959 & 16.40 &  0.15 & 1.903 $\pm$ 0.617 & 1.893 & 0.162 (+0.138/-0.075) & 0.167 & 2.00 $\pm$ 0.50 &   6 &   6 & 50.85 & 58374.7789\\
  90416 & 18.60 &  0.15 & 2.001 $\pm$ 0.703 & 1.797 & 0.041 (+0.034/-0.018) & 0.050 & 2.00 $\pm$ 0.50 &   3 &   3 & 75.85 & 57071.5748\\
  96590 & 16.20 &  0.15 & 1.737 $\pm$ 0.446 & 1.583 & 0.159 (+0.092/-0.058) & 0.192 & 2.00 $\pm$ 0.50 &   4 &   4 & 73.58 & 58167.5872\\
  99907 & 17.90 &  0.15 & 0.734 $\pm$ 0.181 & 0.714 & 0.243 (+0.141/-0.089) & 0.263 & 2.00 $\pm$ 0.50 &   5 &   4 & 60.12 & 58094.0256\\
  A3067 & 16.80 &  0.15 & 1.342 $\pm$ 0.323 & 1.266 & 0.187 (+0.161/-0.087) & 0.214 & 2.00 $\pm$ 0.50 &   6 &   6 & 67.36 & 57069.8124\\
  D6793 & 18.30 &  0.15 & 0.998 $\pm$ 0.270 & 0.984 & 0.186 (+0.143/-0.081) & 0.195 & 2.00 $\pm$ 0.50 &  10 &  10 & 55.22 & 57880.0746\\
  D6839 & 18.70 &  0.15 & 0.611 $\pm$ 0.157 & 0.574 & 0.157 (+0.144/-0.075) & 0.180 & 2.00 $\pm$ 0.50 &   0 &   1 & 68.00 & 58126.7572\\
  D6923 & 16.20 &  0.15 & 3.042 $\pm$ 0.946 & 3.066 & 0.097 (+0.112/-0.052) & 0.096 & 2.00 $\pm$ 0.50 &   4 &   4 & 37.95 & 57951.2161\\
  D7032 & 16.60 &  0.15 & 2.046 $\pm$ 0.552 & 2.040 & 0.120 (+0.083/-0.049) & 0.123 & 2.00 $\pm$ 0.50 &   1 &   1 & 49.54 & 57701.1332\\
  D7078 & 16.20 &  0.15 & 1.524 $\pm$ 0.304 & 1.455 & 0.204 (+0.090/-0.062) & 0.229 & 2.00 $\pm$ 0.50 &   5 &   5 & 64.88 & 57904.7839\\
  D7158 & 17.90 &  0.15 & 0.849 $\pm$ 0.167 & 0.776 & 0.187 (+0.099/-0.065) & 0.225 & 2.00 $\pm$ 0.50 &   5 &   5 & 73.11 & 57193.0565\\
  D7805 & 16.60 &  0.15 & 4.738 $\pm$ 1.497 & 4.546 & 0.033 (+0.024/-0.014) & 0.036 & 2.00 $\pm$ 0.50 &   5 &   5 & 63.67 & 58528.2073\\
  D7925 & 15.90 &  0.15 & 2.040 $\pm$ 0.495 & 1.914 & 0.162 (+0.089/-0.058) & 0.187 & 2.00 $\pm$ 0.50 &   3 &   3 & 68.58 & 58185.7849\\
  D8325 & 16.80 &  0.15 & 1.031 $\pm$ 0.249 & 0.961 & 0.240 (+0.130/-0.084) & 0.279 & 2.00 $\pm$ 0.50 &   6 &   6 & 69.68 & 57511.6834\\
  D8404 & 19.10 &  0.15 & 0.477 $\pm$ 0.090 & 0.392 & 0.131 (+0.060/-0.041) & 0.187 & 2.00 $\pm$ 0.50 &   2 &   3 & 86.50 & 57825.5758\\
  D8846 & 16.00 &  0.15 & 2.707 $\pm$ 0.734 & 2.401 & 0.156 (+0.096/-0.059) & 0.197 & 2.00 $\pm$ 0.50 &   6 &   6 & 77.55 & 57881.6361\\
  D8859 & 18.90 &  0.15 & 0.568 $\pm$ 0.175 & 0.514 & 0.191 (+0.215/-0.101) & 0.233 & 2.00 $\pm$ 0.50 &   4 &   4 & 74.51 & 58495.2781\\
  D9345 & 16.70 &  0.15 & 1.620 $\pm$ 0.388 & 1.489 & 0.290 (+0.223/-0.126) & 0.345 & 2.00 $\pm$ 0.50 &   3 &   4 & 72.14 & 58258.4646\\
  E0158 & 18.40 &  0.15 & 0.637 $\pm$ 0.141 & 0.603 & 0.155 (+0.079/-0.052) & 0.176 & 2.00 $\pm$ 0.50 &   4 &   4 & 66.56 & 58108.6917\\
  E0333 & 19.00 &  0.15 & 0.430 $\pm$ 0.108 & 0.381 & 0.287 (+0.194/-0.116) & 0.362 & 2.00 $\pm$ 0.50 &   5 &   5 & 77.56 & 58356.2467\\
  E1484 & 16.40 &  0.15 & 1.287 $\pm$ 0.292 & 1.253 & 0.294 (+0.182/-0.112) & 0.317 & 2.00 $\pm$ 0.50 &   0 &   1 & 59.71 & 57065.4061\\
  E1495 & 18.30 &  0.15 & 0.540 $\pm$ 0.135 & 0.476 & 0.245 (+0.157/-0.096) & 0.313 & 2.00 $\pm$ 0.50 &   3 &   3 & 78.35 & 58603.3027\\
  E1527 & 18.90 &  0.15 & 0.534 $\pm$ 0.153 & 0.466 & 0.167 (+0.135/-0.075) & 0.217 & 2.00 $\pm$ 0.50 &   4 &   4 & 79.86 & 58349.8840\\
  E1851 & 17.70 &  0.15 & 0.893 $\pm$ 0.186 & 0.843 & 0.242 (+0.126/-0.083) & 0.277 & 2.00 $\pm$ 0.50 &   7 &   7 & 67.34 & 57664.8948\\
  E3947 & 15.30 &  0.15 & 2.945 $\pm$ 0.760 & 2.870 & 0.139 (+0.114/-0.063) & 0.150 & 2.00 $\pm$ 0.50 &  19 &  20 & 59.43 & 58623.9676\\
  E4332 & 16.50 &  0.15 & 1.770 $\pm$ 0.437 & 1.529 & 0.144 (+0.090/-0.055) & 0.190 & 2.00 $\pm$ 0.50 &  11 &  12 & 80.98 & 58369.3077\\
  E4898 & 18.80 &  0.15 & 0.383 $\pm$ 0.092 & 0.377 & 0.309 (+0.212/-0.126) & 0.327 & 2.00 $\pm$ 0.50 &   5 &   5 & 56.19 & 58608.4373\\
  F2671 & 20.10 &  0.15 & 0.326 $\pm$ 0.068 & 0.311 & 0.216 (+0.100/-0.068) & 0.242 & 2.00 $\pm$ 0.50 &   6 &   6 & 64.72 & 57545.9042\\
  F2685 & 19.20 &  0.15 & 0.397 $\pm$ 0.091 & 0.331 & 0.170 (+0.087/-0.057) & 0.238 & 2.00 $\pm$ 0.50 &   3 &   3 & 85.09 & 57710.6282\\
  F3953 & 16.80 &  0.15 & 1.263 $\pm$ 0.229 & 1.211 & 0.164 (+0.065/-0.047) & 0.182 & 2.00 $\pm$ 0.50 &   6 &   7 & 63.82 & 57432.4794\\
  F4229 & 16.50 &  0.15 & 1.339 $\pm$ 0.387 & 1.160 & 0.226 (+0.229/-0.114) & 0.297 & 2.00 $\pm$ 0.50 &   1 &   2 & 80.72 & 56870.7552\\
  F4275 & 20.10 &  0.15 & 0.262 $\pm$ 0.054 & 0.215 & 0.193 (+0.141/-0.081) & 0.276 & 2.00 $\pm$ 0.50 &   1 &   1 & 86.37 & 56819.3343\\
  F4555 & 16.70 &  0.15 & 1.625 $\pm$ 0.450 & 1.564 & 0.122 (+0.077/-0.047) & 0.134 & 2.00 $\pm$ 0.50 &   6 &   6 & 62.90 & 57513.8059\\
  F5140 & 17.40 &  0.15 & 1.629 $\pm$ 0.406 & 1.453 & 0.054 (+0.036/-0.022) & 0.068 & 2.00 $\pm$ 0.50 &   3 &   3 & 76.80 & 58383.3960\\
  F9504 & 17.00 &  0.15 & 3.075 $\pm$ 0.674 & 2.800 & 0.057 (+0.028/-0.019) & 0.069 & 2.00 $\pm$ 0.50 &   7 &   7 & 73.77 & 57324.0052\\
  G1989 & 17.20 &  0.15 & 1.137 $\pm$ 0.244 & 1.089 & 0.229 (+0.172/-0.098) & 0.255 & 2.00 $\pm$ 0.50 &   2 &   2 & 64.07 & 57671.2221\\
  G2058 & 17.90 &  0.15 & 1.608 $\pm$ 0.466 & 1.571 & 0.054 (+0.077/-0.032) & 0.058 & 2.00 $\pm$ 0.50 &   1 &   1 & 58.57 & 56856.9883\\
  G2142 & 18.70 &  0.15 & 0.724 $\pm$ 0.209 & 0.672 & 0.170 (+0.137/-0.076) & 0.200 & 2.00 $\pm$ 0.50 &  11 &  11 & 70.64 & 57645.3273\\
  G2269 & 17.00 &  0.15 & 1.196 $\pm$ 0.339 & 1.138 & 0.187 (+0.134/-0.078) & 0.210 & 2.00 $\pm$ 0.50 &   3 &   3 & 65.50 & 57682.5659\\
  G2361 & 20.10 &  0.15 & 0.213 $\pm$ 0.044 & 0.203 & 0.285 (+0.129/-0.089) & 0.320 & 2.00 $\pm$ 0.50 &   3 &   3 & 65.10 & 58563.1065\\
  G2385 & 18.60 &  0.15 & 0.652 $\pm$ 0.168 & 0.577 & 0.188 (+0.110/-0.069) & 0.239 & 2.00 $\pm$ 0.50 &  11 &  10 & 77.94 & 57432.1497\\
  G2474 & 18.40 &  0.15 & 0.558 $\pm$ 0.120 & 0.509 & 0.247 (+0.188/-0.107) & 0.299 & 2.00 $\pm$ 0.50 &   0 &   5 & 73.46 & 58285.3802\\
  G2687 & 19.00 &  0.15 & 0.537 $\pm$ 0.137 & 0.450 & 0.177 (+0.116/-0.070) & 0.246 & 2.00 $\pm$ 0.50 &   4 &   4 & 84.67 & 58444.6468\\
  G2882 & 18.70 &  0.15 & 0.613 $\pm$ 0.144 & 0.594 & 0.133 (+0.070/-0.046) & 0.146 & 2.00 $\pm$ 0.50 &   4 &   4 & 61.12 & 58172.3197\\
  G2911 & 19.20 &  0.15 & 0.392 $\pm$ 0.079 & 0.331 & 0.189 (+0.104/-0.067) & 0.258 & 2.00 $\pm$ 0.50 &   5 &   5 & 83.60 & 57701.6729\\
  G3081 & 18.40 &  0.15 & 0.537 $\pm$ 0.118 & 0.456 & 0.228 (+0.122/-0.079) & 0.309 & 2.00 $\pm$ 0.50 &   7 &   7 & 82.97 & 58558.8293\\
  G3243 & 16.50 &  0.15 & 1.362 $\pm$ 0.375 & 1.284 & 0.204 (+0.146/-0.085) & 0.234 & 2.00 $\pm$ 0.50 &   9 &   9 & 67.51 & 58170.1108\\
  I9173 & 18.90 &  0.15 & 1.222 $\pm$ 0.282 & 1.184 & 0.032 (+0.017/-0.011) & 0.035 & 2.00 $\pm$ 0.50 &   4 &   4 & 61.11 & 58131.2692\\
  J4268 & 18.20 &  0.15 & 2.362 $\pm$ 0.579 & 2.039 & 0.038 (+0.021/-0.013) & 0.050 & 2.00 $\pm$ 0.50 &   2 &   3 & 81.08 & 57314.9369\\
  J4386 & 16.90 &  0.15 & 2.329 $\pm$ 0.602 & 1.988 & 0.118 (+0.069/-0.043) & 0.158 & 2.00 $\pm$ 0.50 &   3 &   2 & 82.38 & 57500.9009\\
  K0754 & 18.60 &  0.15 & 0.496 $\pm$ 0.123 & 0.462 & 0.261 (+0.167/-0.102) & 0.304 & 2.00 $\pm$ 0.50 &   0 &   8 & 69.59 & 57340.0457\\
  K4131 & 19.80 &  0.15 & 0.225 $\pm$ 0.057 & 0.189 & 0.303 (+0.194/-0.118) & 0.417 & 2.00 $\pm$ 0.50 &   1 &   1 & 84.10 & 57006.3744\\
  K4232 & 18.30 &  0.15 & 0.693 $\pm$ 0.117 & 0.638 & 0.310 (+0.115/-0.084) & 0.369 & 2.00 $\pm$ 0.50 &   7 &   7 & 71.93 & 57345.2830\\
  K8565 & 17.10 &  0.15 & 1.240 $\pm$ 0.371 & 1.216 & 0.166 (+0.121/-0.070) & 0.177 & 2.00 $\pm$ 0.50 &   0 &   7 & 57.21 & 57453.5533\\
  L5442 & 18.50 &  0.15 & 0.717 $\pm$ 0.198 & 0.700 & 0.150 (+0.094/-0.058) & 0.161 & 2.00 $\pm$ 0.50 &  13 &  12 & 59.04 & 57599.9503\\
  L9021 & 16.90 &  0.15 & 1.754 $\pm$ 0.547 & 1.686 & 0.141 (+0.102/-0.059) & 0.156 & 2.00 $\pm$ 0.50 &   7 &   7 & 63.23 & 57537.7883\\
  M0839 & 17.30 &  0.15 & 1.771 $\pm$ 0.525 & 1.724 & 0.122 (+0.083/-0.049) & 0.132 & 2.00 $\pm$ 0.50 &   4 &   4 & 59.82 & 58257.9219\\
  M0909 & 18.20 &  0.15 & 2.377 $\pm$ 0.719 & 2.194 & 0.042 (+0.030/-0.017) & 0.050 & 2.00 $\pm$ 0.50 &   2 &   2 & 71.48 & 58008.2539\\
  N5756 & 18.70 &  0.15 & 0.593 $\pm$ 0.135 & 0.537 & 0.193 (+0.098/-0.065) & 0.236 & 2.00 $\pm$ 0.50 &   3 &   5 & 74.48 & 57130.4819\\
  N9849 & 18.60 &  0.15 & 0.609 $\pm$ 0.116 & 0.564 & 0.308 (+0.140/-0.096) & 0.363 & 2.00 $\pm$ 0.50 &   9 &   9 & 70.88 & 57709.5789\\
  O2191 & 19.00 &  0.15 & 0.384 $\pm$ 0.104 & 0.329 & 0.294 (+0.180/-0.112) & 0.394 & 2.00 $\pm$ 0.50 &   3 &   3 & 82.16 & 57217.3365\\
  O2643 & 17.40 &  0.15 & 2.717 $\pm$ 0.831 & 2.382 & 0.027 (+0.019/-0.011) & 0.035 & 2.00 $\pm$ 0.50 &   2 &   2 & 79.13 & 57236.6518\\
  O2708 & 18.10 &  0.15 & 0.703 $\pm$ 0.167 & 0.610 & 0.179 (+0.119/-0.072) & 0.234 & 2.00 $\pm$ 0.50 &   5 &   5 & 80.44 & 58244.4696\\
  O4670 & 18.50 &  0.15 & 0.931 $\pm$ 0.232 & 0.902 & 0.121 (+0.068/-0.043) & 0.132 & 2.00 $\pm$ 0.50 &  14 &  14 & 61.11 & 58623.4258\\
  O7360 & 19.20 &  0.15 & 0.606 $\pm$ 0.174 & 0.566 & 0.100 (+0.116/-0.054) & 0.117 & 2.00 $\pm$ 0.50 &   0 &   7 & 69.44 & 56873.8031\\
  O8083 & 16.00 &  0.15 & 4.960 $\pm$ 1.446 & 4.960 & 0.047 (+0.032/-0.019) & 0.048 & 2.00 $\pm$ 0.50 &   1 &   1 & 47.56 & 57759.6806\\
  P0697 & 17.80 &  0.15 & 1.119 $\pm$ 0.339 & 0.976 & 0.152 (+0.106/-0.062) & 0.197 & 2.00 $\pm$ 0.50 &   7 &   7 & 79.76 & 58590.4201\\
  P1346 & 16.80 &  0.15 & 1.250 $\pm$ 0.256 & 1.093 & 0.229 (+0.103/-0.071) & 0.296 & 2.00 $\pm$ 0.50 &   9 &  10 & 79.44 & 56708.3470\\
  P2558 & 20.20 &  0.15 & 0.890 $\pm$ 0.211 & 0.816 & 0.041 (+0.022/-0.014) & 0.049 & 2.00 $\pm$ 0.50 &   4 &   4 & 72.57 & 57806.9828\\
  P9221 & 19.10 &  0.15 & 0.574 $\pm$ 0.158 & 0.531 & 0.148 (+0.093/-0.057) & 0.175 & 2.00 $\pm$ 0.50 &   2 &   2 & 71.29 & 56738.1250\\
  Q2623 & 18.60 &  0.15 & 0.543 $\pm$ 0.127 & 0.514 & 0.218 (+0.122/-0.078) & 0.247 & 2.00 $\pm$ 0.50 &   0 &  11 & 66.62 & 58131.8596\\
  Q5196 & 18.80 &  0.15 & 0.431 $\pm$ 0.103 & 0.381 & 0.267 (+0.143/-0.093) & 0.340 & 2.00 $\pm$ 0.50 &   9 &   9 & 78.14 & 58133.2761\\
  Q7223 & 18.10 &  0.15 & 1.156 $\pm$ 0.283 & 1.077 & 0.145 (+0.080/-0.052) & 0.169 & 2.00 $\pm$ 0.50 &   3 &   3 & 69.92 & 57404.5682\\
  R5974 & 18.50 &  0.15 & 0.587 $\pm$ 0.184 & 0.509 & 0.180 (+0.182/-0.091) & 0.236 & 2.00 $\pm$ 0.50 &   2 &   2 & 80.41 & 57472.5974\\
  R7142 & 16.70 &  0.15 & 1.117 $\pm$ 0.337 & 1.054 & 0.208 (+0.170/-0.094) & 0.238 & 2.00 $\pm$ 0.50 &   1 &   1 & 67.45 & 57109.0293\\
  R7475 & 20.10 &  0.15 & 0.306 $\pm$ 0.075 & 0.259 & 0.186 (+0.147/-0.082) & 0.254 & 2.00 $\pm$ 0.50 &   3 &   3 & 83.69 & 56883.8944\\
  R9816 & 19.10 &  0.15 & 0.461 $\pm$ 0.114 & 0.413 & 0.162 (+0.109/-0.065) & 0.201 & 2.00 $\pm$ 0.50 &   7 &   7 & 76.21 & 58241.4700\\
  S1365 & 17.40 &  0.15 & 0.976 $\pm$ 0.188 & 0.916 & 0.206 (+0.112/-0.073) & 0.237 & 2.00 $\pm$ 0.50 &   6 &   6 & 68.42 & 58106.5563\\
  S6079 & 19.20 &  0.15 & 0.449 $\pm$ 0.105 & 0.406 & 0.205 (+0.107/-0.070) & 0.251 & 2.00 $\pm$ 0.50 &   4 &   4 & 74.87 & 56945.3829\\
  U2311 & 19.40 &  0.15 & 0.374 $\pm$ 0.086 & 0.320 & 0.187 (+0.112/-0.070) & 0.250 & 2.00 $\pm$ 0.50 &   2 &   2 & 81.97 & 58103.1528\\
  U3250 & 18.10 &  0.15 & 0.592 $\pm$ 0.159 & 0.537 & 0.258 (+0.209/-0.115) & 0.314 & 2.00 $\pm$ 0.50 &   4 &   4 & 74.39 & 57236.3529\\
  U6383 & 21.50 &  0.15 & 0.135 $\pm$ 0.023 & 0.110 & 0.184 (+0.069/-0.050) & 0.267 & 2.00 $\pm$ 0.50 &   2 &   2 & 87.50 & 58139.6795\\
  U6462 & 18.50 &  0.15 & 0.909 $\pm$ 0.195 & 0.894 & 0.204 (+0.140/-0.083) & 0.216 & 2.00 $\pm$ 0.50 &  11 &  11 & 56.06 & 57031.9860\\
  U8635 & 21.90 & -0.12 & 0.339 $\pm$ 0.102 & 0.290 & 0.027 (+0.061/-0.019) & 0.036 & 2.00 $\pm$ 0.50 &   3 &   3 & 82.30 & 57694.4485\\
  V0560 & 17.80 &  0.15 & 0.852 $\pm$ 0.252 & 0.778 & 0.176 (+0.135/-0.077) & 0.212 & 2.00 $\pm$ 0.50 &   5 &   5 & 73.42 & 58016.7090\\
  V7255 & 19.80 &  0.15 & 0.406 $\pm$ 0.090 & 0.396 & 0.200 (+0.098/-0.066) & 0.216 & 2.00 $\pm$ 0.50 &   4 &   4 & 59.44 & 57844.5278\\
  V8411 & 17.10 &  0.15 & 1.112 $\pm$ 0.246 & 1.022 & 0.242 (+0.142/-0.089) & 0.289 & 2.00 $\pm$ 0.50 &   9 &   9 & 72.23 & 58571.1116\\
  V9988 & 19.50 &  0.15 & 0.423 $\pm$ 0.094 & 0.376 & 0.139 (+0.069/-0.046) & 0.175 & 2.00 $\pm$ 0.50 &   5 &   7 & 77.06 & 57833.5004\\
  W5102 & 20.10 &  0.15 & 0.269 $\pm$ 0.055 & 0.229 & 0.410 (+0.230/-0.147) & 0.552 & 2.00 $\pm$ 0.50 &   7 &   9 & 82.51 & 57097.4579\\
  W6388 & 18.30 &  0.15 & 2.027 $\pm$ 0.498 & 1.826 & 0.023 (+0.013/-0.008) & 0.029 & 2.00 $\pm$ 0.50 &   3 &   3 & 75.37 & 58036.9963\\
  W9770 & 17.90 &  0.15 & 0.952 $\pm$ 0.157 & 0.913 & 0.183 (+0.094/-0.062) & 0.203 & 2.00 $\pm$ 0.50 &  13 &  13 & 63.72 & 57170.1513\\
  X7228 & 15.40 &  0.15 & 3.443 $\pm$ 0.806 & 3.451 & 0.107 (+0.067/-0.041) & 0.108 & 2.00 $\pm$ 0.50 &   1 &   1 & 45.67 & 58097.5589\\
  Y4074 & 18.70 &  0.15 & 0.507 $\pm$ 0.118 & 0.463 & 0.223 (+0.115/-0.076) & 0.269 & 2.00 $\pm$ 0.50 &   6 &   6 & 73.28 & 57726.7940\\
  Y9074 & 18.30 &  0.15 & 1.782 $\pm$ 0.472 & 1.701 & 0.039 (+0.023/-0.014) & 0.043 & 2.00 $\pm$ 0.50 &   6 &   6 & 64.97 & 57677.3633\\
  Z0988 & 16.80 &  0.15 & 1.576 $\pm$ 0.524 & 1.560 & 0.143 (+0.111/-0.063) & 0.150 & 2.00 $\pm$ 0.50 &   1 &   1 & 53.54 & 58186.3185\\
  Z7024 & 19.50 &  0.15 & 0.397 $\pm$ 0.106 & 0.328 & 0.179 (+0.109/-0.068) & 0.256 & 2.00 $\pm$ 0.50 &   8 &   8 & 86.21 & 57641.5206\\
  Z7028 & 19.40 &  0.15 & 0.409 $\pm$ 0.098 & 0.389 & 0.183 (+0.109/-0.068) & 0.207 & 2.00 $\pm$ 0.50 &   0 &   3 & 65.92 & 57708.8640\\
  Z9369 & 18.50 &  0.15 & 0.741 $\pm$ 0.167 & 0.702 & 0.134 (+0.067/-0.045) & 0.152 & 2.00 $\pm$ 0.50 &   5 &   5 & 66.51 & 57596.0220\\
  a3599 & 21.00 &  0.15 & 0.197 $\pm$ 0.041 & 0.161 & 0.212 (+0.099/-0.067) & 0.307 & 2.00 $\pm$ 0.50 &   1 &   1 & 87.17 & 58219.8183\\
  a3831 & 19.20 &  0.15 & 0.443 $\pm$ 0.132 & 0.376 & 0.229 (+0.156/-0.093) & 0.311 & 2.00 $\pm$ 0.50 &   4 &   4 & 83.19 & 57677.4248\\
  a4877 & 17.30 &  0.15 & 1.801 $\pm$ 0.402 & 1.755 & 0.117 (+0.058/-0.039) & 0.126 & 2.00 $\pm$ 0.50 &   5 &   5 & 59.53 & 58533.3694\\
  a5246 & 15.90 &  0.15 & 3.472 $\pm$ 1.076 & 3.397 & 0.096 (+0.081/-0.044) & 0.103 & 2.00 $\pm$ 0.50 &   8 &   9 & 58.13 & 56772.9552\\
  a7684 & 18.20 &  0.15 & 0.930 $\pm$ 0.249 & 0.900 & 0.115 (+0.070/-0.044) & 0.126 & 2.00 $\pm$ 0.50 &   5 &   6 & 61.45 & 58576.8986\\
  a9264 & 16.30 &  0.15 & 1.957 $\pm$ 0.566 & 1.722 & 0.129 (+0.085/-0.051) & 0.164 & 2.00 $\pm$ 0.50 &  14 &  14 & 78.64 & 57711.9925\\
  a9264 & 16.30 &  0.15 & 1.856 $\pm$ 0.388 & 1.834 & 0.157 (+0.079/-0.053) & 0.165 & 2.00 $\pm$ 0.50 &   3 &   3 & 54.36 & 57683.8577\\
  b0702 & 17.40 &  0.15 & 1.410 $\pm$ 0.365 & 1.399 & 0.113 (+0.094/-0.051) & 0.117 & 2.00 $\pm$ 0.50 &   1 &   1 & 52.55 & 57877.3592\\
  b1336 & 17.40 &  0.15 & 1.171 $\pm$ 0.307 & 1.125 & 0.205 (+0.122/-0.076) & 0.227 & 2.00 $\pm$ 0.50 &   7 &   7 & 63.44 & 58055.0627\\
  b7097 & 19.60 &  0.15 & 0.376 $\pm$ 0.102 & 0.326 & 0.224 (+0.180/-0.100) & 0.294 & 2.00 $\pm$ 0.50 &   4 &   4 & 80.61 & 56667.4645\\
  c0981 & 18.70 &  0.15 & 0.496 $\pm$ 0.114 & 0.449 & 0.238 (+0.141/-0.088) & 0.291 & 2.00 $\pm$ 0.50 &   0 &   1 & 74.77 & 57182.5795\\
  c5252 & 19.30 &  0.15 & 0.360 $\pm$ 0.096 & 0.315 & 0.260 (+0.179/-0.106) & 0.335 & 2.00 $\pm$ 0.50 &   0 &  10 & 79.15 & 56720.9055\\
  c5843 & 20.50 &  0.15 & 0.343 $\pm$ 0.079 & 0.305 & 0.095 (+0.057/-0.036) & 0.120 & 2.00 $\pm$ 0.50 &   0 &   5 & 77.47 & 56681.5244\\
  c7668 & 20.40 &  0.15 & 0.285 $\pm$ 0.087 & 0.244 & 0.150 (+0.106/-0.062) & 0.202 & 2.00 $\pm$ 0.50 &   0 &   1 & 82.45 & 56732.0267\\
  c7793 & 16.50 &  0.15 & 1.836 $\pm$ 0.402 & 1.774 & 0.122 (+0.059/-0.040) & 0.133 & 2.00 $\pm$ 0.50 &   6 &   6 & 61.94 & 56934.4322\\
  c9694 & 18.10 &  0.15 & 0.749 $\pm$ 0.204 & 0.657 & 0.177 (+0.127/-0.074) & 0.228 & 2.00 $\pm$ 0.50 &   4 &   4 & 79.08 & 57321.7359\\
  d1508 & 19.10 &  0.15 & 0.393 $\pm$ 0.099 & 0.369 & 0.239 (+0.135/-0.086) & 0.275 & 2.00 $\pm$ 0.50 &   4 &   4 & 68.62 & 57852.2973\\
  d7847 & 18.80 &  0.15 & 0.471 $\pm$ 0.105 & 0.411 & 0.226 (+0.188/-0.102) & 0.292 & 2.00 $\pm$ 0.50 &   5 &   5 & 79.63 & 56805.0624\\
  d8188 & 19.40 &  0.15 & 0.471 $\pm$ 0.127 & 0.451 & 0.188 (+0.115/-0.071) & 0.210 & 2.00 $\pm$ 0.50 &   3 &   3 & 64.19 & 58341.4144\\
  e1856 & 20.20 &  0.15 & 0.975 $\pm$ 0.226 & 0.874 & 0.016 (+0.014/-0.007) & 0.020 & 2.00 $\pm$ 0.50 &   7 &   7 & 76.10 & 56786.8017\\
  e5212 & 18.00 &  0.15 & 0.700 $\pm$ 0.186 & 0.679 & 0.267 (+0.213/-0.118) & 0.290 & 2.00 $\pm$ 0.50 &   5 &   5 & 60.45 & 58274.6650\\
  f3260 & 19.40 &  0.15 & 0.395 $\pm$ 0.090 & 0.347 & 0.196 (+0.123/-0.076) & 0.251 & 2.00 $\pm$ 0.50 &  10 &  13 & 78.51 & 58365.5445\\
  f4287 & 17.70 &  0.15 & 2.450 $\pm$ 0.698 & 2.199 & 0.122 (+0.079/-0.048) & 0.151 & 2.00 $\pm$ 0.50 &   5 &   5 & 75.90 & 56929.1791\\
  f7581 & 17.20 &  0.15 & 1.126 $\pm$ 0.238 & 1.041 & 0.145 (+0.068/-0.046) & 0.171 & 2.00 $\pm$ 0.50 &   4 &   4 & 71.14 & 57795.1005\\
  f7874 & 18.10 &  0.15 & 0.852 $\pm$ 0.210 & 0.812 & 0.156 (+0.096/-0.059) & 0.175 & 2.00 $\pm$ 0.50 &   6 &   6 & 65.30 & 56997.7791\\
  f8094 & 19.60 &  0.15 & 0.972 $\pm$ 0.228 & 0.837 & 0.054 (+0.028/-0.018) & 0.071 & 2.00 $\pm$ 0.50 &   4 &   3 & 81.34 & 57909.3293\\
  f8233 & 18.30 &  0.15 & 1.204 $\pm$ 0.290 & 1.118 & 0.181 (+0.098/-0.063) & 0.212 & 2.00 $\pm$ 0.50 &   8 &  11 & 70.40 & 57069.7785\\
  f8416 & 20.30 &  0.15 & 0.280 $\pm$ 0.071 & 0.245 & 0.171 (+0.098/-0.062) & 0.221 & 2.00 $\pm$ 0.50 &   0 &   3 & 79.41 & 56858.8357\\
  f8929 & 17.00 &  0.15 & 1.809 $\pm$ 0.491 & 1.703 & 0.102 (+0.063/-0.039) & 0.117 & 2.00 $\pm$ 0.50 &   1 &   1 & 67.91 & 58439.6598\\
  f9880 & 19.50 &  0.15 & 0.965 $\pm$ 0.281 & 0.918 & 0.040 (+0.026/-0.016) & 0.045 & 2.00 $\pm$ 0.50 &   2 &   2 & 65.85 & 58510.1131\\
  g0591 & 19.30 &  0.15 & 0.360 $\pm$ 0.088 & 0.313 & 0.277 (+0.163/-0.103) & 0.361 & 2.00 $\pm$ 0.50 &   6 &   6 & 80.04 & 57702.7559\\
  g4965 & 18.80 &  0.15 & 0.632 $\pm$ 0.167 & 0.565 & 0.160 (+0.096/-0.060) & 0.199 & 2.00 $\pm$ 0.50 &  11 &  12 & 76.38 & 57064.3201\\
  g9094 & 20.10 &  0.15 & 0.311 $\pm$ 0.074 & 0.263 & 0.209 (+0.150/-0.088) & 0.285 & 2.00 $\pm$ 0.50 &   8 &   8 & 83.23 & 57116.3349\\
  h1107 & 17.70 &  0.15 & 2.108 $\pm$ 0.494 & 2.098 & 0.033 (+0.020/-0.012) & 0.034 & 2.00 $\pm$ 0.50 &   0 &   3 & 50.79 & 57198.4014\\
  h3953 & 20.70 &  0.15 & 0.332 $\pm$ 0.103 & 0.270 & 0.074 (+0.053/-0.031) & 0.108 & 2.00 $\pm$ 0.50 &   3 &   3 & 87.55 & 57705.3136\\
  h6037 & 18.60 &  0.15 & 0.683 $\pm$ 0.211 & 0.645 & 0.159 (+0.191/-0.087) & 0.182 & 2.00 $\pm$ 0.50 &   5 &   5 & 67.43 & 57020.8262\\
  h6325 & 18.40 &  0.15 & 0.709 $\pm$ 0.171 & 0.664 & 0.157 (+0.085/-0.055) & 0.182 & 2.00 $\pm$ 0.50 &   3 &   3 & 68.77 & 58140.3297\\
  h6329 & 20.10 &  0.15 & 1.238 $\pm$ 0.293 & 1.128 & 0.053 (+0.028/-0.018) & 0.064 & 2.00 $\pm$ 0.50 &   7 &   7 & 73.70 & 57147.8066\\
  h6724 & 19.90 &  0.15 & 0.422 $\pm$ 0.113 & 0.341 & 0.096 (+0.074/-0.042) & 0.142 & 2.00 $\pm$ 0.50 &   2 &   4 & 88.32 & 57231.4184\\
  h6775 & 16.50 &  0.15 & 1.372 $\pm$ 0.393 & 1.224 & 0.180 (+0.122/-0.073) & 0.226 & 2.00 $\pm$ 0.50 &   2 &   2 & 76.84 & 57134.9233\\
  h8661 & 19.40 &  0.15 & 0.331 $\pm$ 0.081 & 0.278 & 0.297 (+0.194/-0.117) & 0.411 & 2.00 $\pm$ 0.50 &   5 &   6 & 84.37 & 57425.7596\\
  h9877 & 19.50 &  0.15 & 0.613 $\pm$ 0.107 & 0.527 & 0.085 (+0.032/-0.023) & 0.113 & 2.00 $\pm$ 0.50 &   5 &   6 & 81.66 & 57177.2392\\
  i4004 & 20.50 &  0.15 & 0.343 $\pm$ 0.111 & 0.298 & 0.117 (+0.136/-0.063) & 0.152 & 2.00 $\pm$ 0.50 &   7 &   8 & 80.37 & 56696.6192\\
  i4193 & 18.10 &  0.15 & 0.722 $\pm$ 0.152 & 0.615 & 0.197 (+0.096/-0.064) & 0.266 & 2.00 $\pm$ 0.50 &   4 &   4 & 82.76 & 58221.0599\\
  i6924 & 20.20 &  0.15 & 0.328 $\pm$ 0.073 & 0.304 & 0.191 (+0.095/-0.063) & 0.225 & 2.00 $\pm$ 0.50 &   5 &   5 & 70.74 & 57367.5552\\
  j0142 & 19.90 &  0.15 & 0.297 $\pm$ 0.064 & 0.254 & 0.339 (+0.198/-0.125) & 0.454 & 2.00 $\pm$ 0.50 &   4 &   4 & 82.02 & 57343.1318\\
  j0160 & 16.70 &  0.15 & 1.898 $\pm$ 0.527 & 1.823 & 0.119 (+0.075/-0.046) & 0.132 & 2.00 $\pm$ 0.50 &   5 &   5 & 63.43 & 57470.2564\\
  j0263 & 18.80 &  0.15 & 0.765 $\pm$ 0.160 & 0.674 & 0.191 (+0.088/-0.060) & 0.243 & 2.00 $\pm$ 0.50 &   7 &   7 & 78.52 & 57411.3131\\
  j0649 & 19.50 &  0.15 & 0.343 $\pm$ 0.087 & 0.295 & 0.217 (+0.123/-0.079) & 0.287 & 2.00 $\pm$ 0.50 &   5 &   7 & 81.35 & 57300.8436\\
  j0894 & 18.30 &  0.15 & 0.898 $\pm$ 0.227 & 0.770 & 0.119 (+0.076/-0.046) & 0.159 & 2.00 $\pm$ 0.50 &   1 &   1 & 81.89 & 58220.2739\\
  j2302 & 20.40 &  0.15 & 0.837 $\pm$ 0.259 & 0.683 & 0.021 (+0.019/-0.010) & 0.031 & 2.00 $\pm$ 0.50 &   1 &   1 & 87.37 & 57378.9721\\
  j2376 & 20.00 &  0.15 & 0.292 $\pm$ 0.063 & 0.251 & 0.301 (+0.153/-0.101) & 0.401 & 2.00 $\pm$ 0.50 &   4 &   4 & 81.59 & 57389.8650\\
  j3563 & 20.20 &  0.15 & 0.210 $\pm$ 0.042 & 0.177 & 0.300 (+0.182/-0.113) & 0.412 & 2.00 $\pm$ 0.50 &   7 &   7 & 83.89 & 57378.4104\\
  j4094 & 21.90 &  0.15 & 0.157 $\pm$ 0.052 & 0.133 & 0.132 (+0.126/-0.064) & 0.180 & 2.00 $\pm$ 0.50 &   5 &   5 & 83.69 & 56694.1089\\
  j5554 & 18.00 &  0.15 & 2.064 $\pm$ 0.525 & 1.831 & 0.037 (+0.021/-0.013) & 0.047 & 2.00 $\pm$ 0.50 &   3 &   3 & 77.49 & 57393.9945\\
  k2775 & 19.40 &  0.15 & 1.391 $\pm$ 0.432 & 1.314 & 0.016 (+0.026/-0.010) & 0.018 & 2.00 $\pm$ 0.50 &   1 &   1 & 67.31 & 57654.5868\\
  k4797 & 18.30 &  0.15 & 1.047 $\pm$ 0.270 & 0.975 & 0.103 (+0.060/-0.038) & 0.120 & 2.00 $\pm$ 0.50 &   4 &   4 & 69.80 & 57601.2832\\
  k5824 & 21.70 &  0.15 & 0.242 $\pm$ 0.053 & 0.223 & 0.076 (+0.037/-0.025) & 0.091 & 2.00 $\pm$ 0.50 &   6 &   6 & 72.18 & 57469.1084\\
  k6508 & 18.60 &  0.15 & 0.822 $\pm$ 0.257 & 0.721 & 0.108 (+0.078/-0.045) & 0.138 & 2.00 $\pm$ 0.50 &  10 &  10 & 78.95 & 56771.4604\\
  k7309 & 19.90 &  0.15 & 0.169 $\pm$ 0.042 & 0.153 & 0.365 (+0.206/-0.132) & 0.448 & 2.00 $\pm$ 0.50 &   2 &   2 & 74.90 & 58289.2730\\
  k7336 & 20.50 &  0.15 & 0.237 $\pm$ 0.051 & 0.218 & 0.234 (+0.112/-0.076) & 0.278 & 2.00 $\pm$ 0.50 &   5 &   5 & 72.00 & 57568.5081\\
  k7963 & 18.60 &  0.15 & 0.673 $\pm$ 0.163 & 0.595 & 0.278 (+0.164/-0.103) & 0.354 & 2.00 $\pm$ 0.50 &   5 &   5 & 78.07 & 57674.1217\\
  k8583 & 17.90 &  0.15 & 0.800 $\pm$ 0.175 & 0.671 & 0.171 (+0.083/-0.056) & 0.236 & 2.00 $\pm$ 0.50 &   4 &   5 & 84.40 & 57707.6874\\
  l1240 & 21.70 &  0.15 & 0.119 $\pm$ 0.024 & 0.097 & 0.283 (+0.175/-0.108) & 0.414 & 2.00 $\pm$ 0.50 &   3 &   4 & 87.77 & 57649.1891\\
  l1241 & 18.20 &  0.15 & 2.598 $\pm$ 0.714 & 2.251 & 0.064 (+0.040/-0.025) & 0.084 & 2.00 $\pm$ 0.50 &   4 &   4 & 80.59 & 57603.1260\\
  l1323 & 19.40 &  0.15 & 1.312 $\pm$ 0.323 & 1.263 & 0.060 (+0.033/-0.022) & 0.067 & 2.00 $\pm$ 0.50 &   6 &   7 & 62.92 & 57570.9236\\
  l1956 & 19.20 &  0.15 & 0.430 $\pm$ 0.110 & 0.406 & 0.304 (+0.175/-0.111) & 0.346 & 2.00 $\pm$ 0.50 &   6 &   6 & 66.88 & 57353.6623\\
  l5534 & 21.30 &  0.15 & 0.194 $\pm$ 0.059 & 0.154 & 0.124 (+0.088/-0.051) & 0.188 & 2.00 $\pm$ 0.50 &   1 &   1 & 89.74 & 58419.8649\\
  l7162 & 20.60 &  0.15 & 0.263 $\pm$ 0.065 & 0.226 & 0.160 (+0.108/-0.065) & 0.213 & 2.00 $\pm$ 0.50 &   2 &   2 & 81.71 & 57629.7126\\
  l7327 & 18.30 &  0.15 & 1.061 $\pm$ 0.317 & 0.921 & 0.117 (+0.080/-0.048) & 0.153 & 2.00 $\pm$ 0.50 &   5 &   5 & 80.31 & 57713.7312\\
  m0004 & 17.20 &  0.15 & 2.077 $\pm$ 0.360 & 1.979 & 0.125 (+0.064/-0.042) & 0.140 & 2.00 $\pm$ 0.50 &  11 &  11 & 65.29 & 57718.9329\\
  m0823 & 20.10 &  0.15 & 0.254 $\pm$ 0.054 & 0.217 & 0.248 (+0.153/-0.095) & 0.332 & 2.00 $\pm$ 0.50 &   2 &   2 & 82.02 & 57753.6418\\
  m2488 & 19.70 &  0.15 & 0.301 $\pm$ 0.076 & 0.255 & 0.257 (+0.194/-0.111) & 0.350 & 2.00 $\pm$ 0.50 &   0 &   2 & 83.30 & 57490.6980\\
  m3563 & 18.70 &  0.15 & 0.639 $\pm$ 0.127 & 0.621 & 0.180 (+0.079/-0.055) & 0.195 & 2.00 $\pm$ 0.50 &   8 &   8 & 60.38 & 57809.0954\\
  m6739 & 19.60 &  0.15 & 0.395 $\pm$ 0.108 & 0.338 & 0.188 (+0.117/-0.072) & 0.252 & 2.00 $\pm$ 0.50 &   1 &   1 & 82.17 & 57755.7736\\
  m8450 & 17.20 &  0.15 & 1.357 $\pm$ 0.419 & 1.327 & 0.127 (+0.090/-0.053) & 0.135 & 2.00 $\pm$ 0.50 &   0 &   2 & 58.14 & 57860.9007\\
  n5323 & 19.60 &  0.15 & 0.968 $\pm$ 0.214 & 0.897 & 0.040 (+0.020/-0.013) & 0.047 & 2.00 $\pm$ 0.50 &   8 &   8 & 70.93 & 56834.5034\\
  n5829 & 18.70 &  0.15 & 0.793 $\pm$ 0.190 & 0.780 & 0.091 (+0.090/-0.045) & 0.097 & 2.00 $\pm$ 0.50 &   3 &   4 & 56.43 & 58322.3134\\
  n6817 & 19.90 &  0.15 & 0.287 $\pm$ 0.077 & 0.248 & 0.160 (+0.125/-0.070) & 0.211 & 2.00 $\pm$ 0.50 &   4 &   4 & 80.92 & 58022.1404\\
  n6861 & 18.00 &  0.15 & 0.857 $\pm$ 0.237 & 0.825 & 0.137 (+0.086/-0.053) & 0.151 & 2.00 $\pm$ 0.50 &   6 &   6 & 62.67 & 57994.1245\\
  n7117 & 18.30 &  0.15 & 0.856 $\pm$ 0.230 & 0.813 & 0.239 (+0.155/-0.094) & 0.270 & 2.00 $\pm$ 0.50 &   6 &   5 & 66.22 & 58171.3180\\
  n8548 & 20.00 &  0.15 & 0.273 $\pm$ 0.063 & 0.251 & 0.245 (+0.146/-0.091) & 0.291 & 2.00 $\pm$ 0.50 &   6 &   6 & 71.93 & 58058.6083\\
  o1647 & 22.30 &  0.15 & 0.128 $\pm$ 0.029 & 0.107 & 0.154 (+0.082/-0.053) & 0.214 & 2.00 $\pm$ 0.50 &   4 &   4 & 84.54 & 56884.2964\\
  o4680 & 19.80 &  0.15 & 0.340 $\pm$ 0.104 & 0.323 & 0.174 (+0.180/-0.088) & 0.196 & 2.00 $\pm$ 0.50 &   6 &   6 & 66.12 & 58096.2616\\
  o4800 & 21.80 &  0.15 & 0.550 $\pm$ 0.166 & 0.464 & 0.020 (+0.016/-0.009) & 0.028 & 2.00 $\pm$ 0.50 &   5 &   5 & 83.64 & 58366.5793\\
  o8772 & 19.80 &  0.15 & 0.277 $\pm$ 0.067 & 0.239 & 0.337 (+0.211/-0.130) & 0.446 & 2.00 $\pm$ 0.50 &   3 &   3 & 81.40 & 58046.9496\\
  o8912 & 19.50 &  0.15 & 0.289 $\pm$ 0.062 & 0.255 & 0.337 (+0.207/-0.128) & 0.429 & 2.00 $\pm$ 0.50 &   4 &   4 & 78.11 & 58171.2531\\
  o8918 & 19.20 &  0.15 & 0.618 $\pm$ 0.199 & 0.571 & 0.115 (+0.086/-0.049) & 0.136 & 2.00 $\pm$ 0.50 &   8 &   8 & 71.50 & 57066.4452\\
  o9520 & 18.90 &  0.15 & 0.667 $\pm$ 0.132 & 0.631 & 0.146 (+0.067/-0.046) & 0.166 & 2.00 $\pm$ 0.50 &   5 &   5 & 67.03 & 58202.2393\\
  o9523 & 19.30 &  0.15 & 1.469 $\pm$ 0.422 & 1.374 & 0.023 (+0.015/-0.009) & 0.027 & 2.00 $\pm$ 0.50 &   2 &   2 & 69.09 & 58100.3090\\
  p1137 & 19.00 &  0.15 & 0.597 $\pm$ 0.171 & 0.529 & 0.178 (+0.175/-0.088) & 0.225 & 2.00 $\pm$ 0.50 &   5 &   5 & 77.83 & 56822.5975\\
  p1684 & 20.70 &  0.15 & 0.191 $\pm$ 0.047 & 0.175 & 0.174 (+0.121/-0.071) & 0.209 & 2.00 $\pm$ 0.50 &   1 &   1 & 73.22 & 58158.8102\\
  p2245 & 19.90 &  0.15 & 0.316 $\pm$ 0.075 & 0.269 & 0.250 (+0.151/-0.094) & 0.337 & 2.00 $\pm$ 0.50 &   3 &   3 & 82.53 & 58088.7794\\
  p3550 & 17.20 &  0.15 & 2.057 $\pm$ 0.780 & 2.012 & 0.064 (+0.064/-0.032) & 0.069 & 2.00 $\pm$ 0.50 &   1 &   1 & 58.18 & 56813.8657\\
  p5010 & 18.90 &  0.15 & 0.570 $\pm$ 0.114 & 0.518 & 0.170 (+0.082/-0.055) & 0.207 & 2.00 $\pm$ 0.50 &   7 &   7 & 74.11 & 58233.6185\\
  p5049 & 19.30 &  0.15 & 0.523 $\pm$ 0.148 & 0.496 & 0.114 (+0.103/-0.054) & 0.129 & 2.00 $\pm$ 0.50 &   4 &   4 & 66.44 & 58217.8786\\
  p6155 & 20.20 &  0.15 & 0.255 $\pm$ 0.059 & 0.221 & 0.238 (+0.175/-0.101) & 0.312 & 2.00 $\pm$ 0.50 &   2 &   2 & 80.63 & 57479.3443\\
  p8507 & 19.80 &  0.15 & 0.416 $\pm$ 0.128 & 0.375 & 0.123 (+0.186/-0.074) & 0.151 & 2.00 $\pm$ 0.50 &   0 &   7 & 75.32 & 58002.9813\\
  p8640 & 20.10 &  0.15 & 0.277 $\pm$ 0.063 & 0.247 & 0.200 (+0.101/-0.067) & 0.252 & 2.00 $\pm$ 0.50 &   5 &   5 & 77.10 & 58379.2253\\
  q0808 & 20.10 &  0.15 & 0.965 $\pm$ 0.262 & 0.882 & 0.027 (+0.021/-0.012) & 0.032 & 2.00 $\pm$ 0.50 &   2 &   3 & 73.12 & 57670.8571\\
  q2684 & 21.10 &  0.15 & 0.140 $\pm$ 0.037 & 0.122 & 0.275 (+0.173/-0.106) & 0.355 & 2.00 $\pm$ 0.50 &   3 &   3 & 79.56 & 58232.2785\\
  q2684 & 21.10 &  0.15 & 0.149 $\pm$ 0.027 & 0.140 & 0.352 (+0.136/-0.098) & 0.407 & 2.00 $\pm$ 0.50 &   5 &   5 & 68.98 & 57873.9345\\
  q3586 & 22.10 &  0.15 & 0.120 $\pm$ 0.036 & 0.099 & 0.164 (+0.163/-0.082) & 0.236 & 2.00 $\pm$ 0.50 &   2 &   2 & 86.64 & 58248.6678\\
  q3625 & 19.60 &  0.15 & 1.251 $\pm$ 0.244 & 1.121 & 0.040 (+0.017/-0.012) & 0.049 & 2.00 $\pm$ 0.50 &   4 &   4 & 76.15 & 58178.0271\\
  q3775 & 19.00 &  0.15 & 0.441 $\pm$ 0.098 & 0.389 & 0.220 (+0.128/-0.081) & 0.281 & 2.00 $\pm$ 0.50 &   4 &   4 & 78.38 & 57090.8911\\
  q3806 & 17.40 &  0.15 & 2.558 $\pm$ 0.674 & 2.555 & 0.030 (+0.022/-0.013) & 0.030 & 2.00 $\pm$ 0.50 &   0 &   1 & 48.58 & 58452.0661\\
  q3815 & 17.50 &  0.15 & 1.083 $\pm$ 0.213 & 1.041 & 0.185 (+0.111/-0.069) & 0.205 & 2.00 $\pm$ 0.50 &   4 &   5 & 63.40 & 58384.7499\\
  q3915 & 18.20 &  0.15 & 0.887 $\pm$ 0.249 & 0.845 & 0.231 (+0.147/-0.090) & 0.259 & 2.00 $\pm$ 0.50 &   5 &   5 & 65.29 & 57011.3303\\
  q4459 & 20.10 &  0.15 & 0.299 $\pm$ 0.078 & 0.289 & 0.203 (+0.120/-0.075) & 0.223 & 2.00 $\pm$ 0.50 &   9 &   9 & 62.01 & 58597.9761\\
  q4530 & 19.50 &  0.15 & 0.371 $\pm$ 0.099 & 0.325 & 0.219 (+0.134/-0.083) & 0.282 & 2.00 $\pm$ 0.50 &   1 &   1 & 79.15 & 58448.2261\\
  q5477 & 19.10 &  0.15 & 0.439 $\pm$ 0.103 & 0.367 & 0.183 (+0.096/-0.063) & 0.254 & 2.00 $\pm$ 0.50 &   2 &   2 & 84.68 & 58568.6464\\
  q8284 & 21.20 &  0.15 & 0.305 $\pm$ 0.081 & 0.288 & 0.064 (+0.039/-0.024) & 0.073 & 2.00 $\pm$ 0.50 &   5 &   6 & 67.66 & 58521.8291\\
  q9450 & 19.30 &  0.15 & 0.657 $\pm$ 0.176 & 0.573 & 0.174 (+0.144/-0.079) & 0.226 & 2.00 $\pm$ 0.50 &   2 &   3 & 79.75 & 58431.7027\\
  q9951 & 21.00 &  0.15 & 0.304 $\pm$ 0.082 & 0.285 & 0.096 (+0.073/-0.042) & 0.111 & 2.00 $\pm$ 0.50 &   3 &   3 & 68.60 & 57850.8881\\
J99T16T & 19.60 &  0.15 & 0.978 $\pm$ 0.210 & 0.849 & 0.052 (+0.025/-0.017) & 0.069 & 2.00 $\pm$ 0.50 &   4 &   4 & 80.39 & 57322.7342\\
K07D41L & 20.70 &  0.15 & 0.201 $\pm$ 0.050 & 0.166 & 0.245 (+0.137/-0.088) & 0.351 & 2.00 $\pm$ 0.50 &   3 &   3 & 86.53 & 56928.7337\\
K07M06K & 20.30 &  0.15 & 0.264 $\pm$ 0.063 & 0.235 & 0.201 (+0.138/-0.082) & 0.253 & 2.00 $\pm$ 0.50 &   2 &   2 & 77.01 & 57555.8048\\
K08G04A & 19.10 &  0.15 & 0.458 $\pm$ 0.109 & 0.409 & 0.290 (+0.157/-0.102) & 0.362 & 2.00 $\pm$ 0.50 &   4 &   4 & 76.70 & 57086.9705\\
K08H01Z & 19.20 &  0.15 & 0.747 $\pm$ 0.233 & 0.700 & 0.071 (+0.051/-0.030) & 0.082 & 2.00 $\pm$ 0.50 &   3 &   3 & 68.80 & 58632.8547\\
K08J30Y & 18.90 &  0.15 & 0.573 $\pm$ 0.112 & 0.535 & 0.165 (+0.096/-0.061) & 0.192 & 2.00 $\pm$ 0.50 &   3 &   3 & 69.71 & 57380.9110\\
K08O02V & 20.30 &  0.15 & 0.294 $\pm$ 0.065 & 0.255 & 0.155 (+0.109/-0.064) & 0.204 & 2.00 $\pm$ 0.50 &   0 &   7 & 80.55 & 58633.3114\\
K08P09R & 22.50 &  0.15 & 0.113 $\pm$ 0.027 & 0.093 & 0.138 (+0.129/-0.067) & 0.197 & 2.00 $\pm$ 0.50 &   0 &   2 & 86.45 & 57507.9383\\
K08S00D & 19.40 &  0.15 & 0.489 $\pm$ 0.123 & 0.441 & 0.122 (+0.069/-0.044) & 0.151 & 2.00 $\pm$ 0.50 &   4 &   4 & 75.47 & 57803.7806\\
K08W00L & 21.70 &  0.15 & 0.697 $\pm$ 0.205 & 0.643 & 0.038 (+0.026/-0.015) & 0.045 & 2.00 $\pm$ 0.50 &   3 &   2 & 71.62 & 58078.4608\\
K09D46L & 22.00 &  0.15 & 0.121 $\pm$ 0.025 & 0.107 & 0.266 (+0.123/-0.084) & 0.339 & 2.00 $\pm$ 0.50 &   5 &   5 & 78.43 & 57525.8599\\
K09U00G & 23.30 &  0.15 & 0.168 $\pm$ 0.042 & 0.136 & 0.049 (+0.028/-0.018) & 0.072 & 2.00 $\pm$ 0.50 &   1 &   1 & 88.15 & 57658.5908\\
K14J25O & 17.80 &  0.15 & 0.732 $\pm$ 0.174 & 0.660 & 0.177 (+0.143/-0.079) & 0.217 & 2.00 $\pm$ 0.50 &   3 &   3 & 75.22 & 56779.2826\\
K16Y08C & 24.60 &  0.15 & 0.045 $\pm$ 0.011 & 0.041 & 0.127 (+0.093/-0.054) & 0.155 & 2.00 $\pm$ 0.50 &   0 &   2 & 75.08 & 57769.3487\\
K17A04M & 19.10 &  0.15 & 0.373 $\pm$ 0.078 & 0.331 & 0.347 (+0.181/-0.119) & 0.439 & 2.00 $\pm$ 0.50 &  12 &  13 & 77.57 & 56861.5611\\
K17A05F & 17.70 &  0.15 & 1.596 $\pm$ 0.349 & 1.525 & 0.078 (+0.038/-0.025) & 0.087 & 2.00 $\pm$ 0.50 &   6 &   6 & 64.66 & 57750.9725\\
K17A19P & 23.60 &  0.15 & 0.049 $\pm$ 0.012 & 0.039 & 0.289 (+0.185/-0.113) & 0.436 & 2.00 $\pm$ 0.50 &   1 &   1 & 89.50 & 57761.8081\\
K17B03M & 23.00 &  0.15 & 0.106 $\pm$ 0.034 & 0.095 & 0.119 (+0.113/-0.058) & 0.147 & 2.00 $\pm$ 0.50 &   2 &   2 & 75.35 & 57794.2338\\
K17B05P & 20.30 &  0.15 & 0.343 $\pm$ 0.073 & 0.321 & 0.134 (+0.067/-0.045) & 0.156 & 2.00 $\pm$ 0.50 &   3 &   3 & 69.44 & 57735.8117\\
K17B06Q & 21.40 &  0.15 & 0.120 $\pm$ 0.027 & 0.103 & 0.130 (+0.103/-0.057) & 0.172 & 2.00 $\pm$ 0.50 &   1 &   1 & 81.29 & 57791.2648\\
K17B31M & 20.70 &  0.15 & 0.216 $\pm$ 0.062 & 0.191 & 0.165 (+0.159/-0.081) & 0.210 & 2.00 $\pm$ 0.50 &   4 &   4 & 78.34 & 57937.3490\\
K17B31P & 19.80 &  0.15 & 0.795 $\pm$ 0.242 & 0.704 & 0.034 (+0.035/-0.017) & 0.042 & 2.00 $\pm$ 0.50 &   3 &   3 & 77.76 & 57808.4781\\
K17C00S & 19.80 &  0.15 & 1.011 $\pm$ 0.266 & 0.922 & 0.032 (+0.019/-0.012) & 0.038 & 2.00 $\pm$ 0.50 &   1 &   2 & 73.36 & 57901.3026\\
K17C00S & 19.80 &  0.15 & 0.916 $\pm$ 0.256 & 0.874 & 0.056 (+0.036/-0.022) & 0.062 & 2.00 $\pm$ 0.50 &  15 &  14 & 65.02 & 57874.4810\\
K17C31X & 25.10 &  0.15 & 0.039 $\pm$ 0.010 & 0.035 & 0.133 (+0.075/-0.048) & 0.166 & 2.00 $\pm$ 0.50 &   2 &   2 & 76.35 & 57787.8763\\
K17D34T & 27.80 &  0.15 & 0.018 $\pm$ 0.005 & 0.014 & 0.043 (+0.047/-0.022) & 0.064 & 2.00 $\pm$ 0.50 &   0 &   1 & 88.95 & 57808.5727\\
K17D34W & 23.20 &  0.15 & 0.078 $\pm$ 0.021 & 0.064 & 0.151 (+0.095/-0.058) & 0.219 & 2.00 $\pm$ 0.50 &   0 &   1 & 87.31 & 57816.2267\\
K17D36C & 22.10 &  0.15 & 0.183 $\pm$ 0.056 & 0.145 & 0.036 (+0.039/-0.019) & 0.054 & 2.00 $\pm$ 0.50 &   1 &   1 & 90.37 & 57801.9627\\
K17E00K & 24.10 &  0.15 & 0.040 $\pm$ 0.010 & 0.032 & 0.254 (+0.140/-0.090) & 0.385 & 2.00 $\pm$ 0.50 &   1 &   1 & 89.70 & 57827.6906\\
K17E02S & 19.40 &  0.15 & 0.653 $\pm$ 0.176 & 0.553 & 0.113 (+0.070/-0.043) & 0.154 & 2.00 $\pm$ 0.50 &   1 &   1 & 83.26 & 57863.9265\\
K17E04L & 25.40 &  0.15 & 0.055 $\pm$ 0.020 & 0.044 & 0.041 (+0.071/-0.026) & 0.061 & 2.00 $\pm$ 0.50 &   0 &   1 & 88.70 & 57817.0770\\
K17E13Q & 19.90 &  0.15 & 0.353 $\pm$ 0.093 & 0.300 & 0.137 (+0.114/-0.062) & 0.186 & 2.00 $\pm$ 0.50 &   3 &   3 & 82.86 & 57857.8568\\
K17E13R & 21.50 &  0.15 & 0.137 $\pm$ 0.036 & 0.114 & 0.333 (+0.221/-0.133) & 0.470 & 2.00 $\pm$ 0.50 &   4 &   4 & 85.70 & 57786.4710\\
K17H01C & 20.80 &  0.15 & 0.376 $\pm$ 0.086 & 0.351 & 0.073 (+0.061/-0.033) & 0.085 & 2.00 $\pm$ 0.50 &   2 &   2 & 69.56 & 57876.7530\\
K17H49P & 22.90 &  0.15 & 0.118 $\pm$ 0.040 & 0.098 & 0.087 (+0.088/-0.044) & 0.124 & 2.00 $\pm$ 0.50 &   0 &   4 & 86.09 & 57904.1436\\
K17K27R & 23.50 &  0.15 & 0.057 $\pm$ 0.010 & 0.051 & 0.352 (+0.237/-0.141) & 0.435 & 2.00 $\pm$ 0.50 &   1 &   1 & 75.74 & 57908.9956\\
K17M01B & 18.80 &  0.15 & 0.491 $\pm$ 0.146 & 0.407 & 0.213 (+0.146/-0.086) & 0.301 & 2.00 $\pm$ 0.50 &   1 &   1 & 85.73 & 57955.8959\\
K17M05B & 22.60 &  0.15 & 0.108 $\pm$ 0.031 & 0.088 & 0.181 (+0.120/-0.072) & 0.265 & 2.00 $\pm$ 0.50 &   3 &   3 & 87.89 & 57944.4070\\
K17N05S & 20.70 &  0.15 & 0.588 $\pm$ 0.164 & 0.551 & 0.033 (+0.022/-0.013) & 0.038 & 2.00 $\pm$ 0.50 &   6 &   6 & 68.59 & 57591.1684\\
K17P26L & 22.20 &  0.15 & 0.174 $\pm$ 0.048 & 0.152 & 0.140 (+0.088/-0.054) & 0.182 & 2.00 $\pm$ 0.50 &   4 &   5 & 79.75 & 57989.6921\\
K17Q18M & 18.60 &  0.15 & 0.845 $\pm$ 0.266 & 0.840 & 0.090 (+0.160/-0.058) & 0.093 & 2.00 $\pm$ 0.50 &   0 &  11 & 51.66 & 57917.7248\\
K17R00L & 18.50 &  0.15 & 0.670 $\pm$ 0.196 & 0.579 & 0.187 (+0.153/-0.084) & 0.246 & 2.00 $\pm$ 0.50 &  24 &  26 & 80.78 & 57981.8972\\
K17S14T & 23.50 &  0.15 & 0.110 $\pm$ 0.028 & 0.095 & 0.067 (+0.141/-0.045) & 0.088 & 2.00 $\pm$ 0.50 &   1 &   1 & 80.67 & 58029.8567\\
K17S17R & 18.80 &  0.15 & 1.027 $\pm$ 0.243 & 0.975 & 0.067 (+0.036/-0.023) & 0.076 & 2.00 $\pm$ 0.50 &   4 &   4 & 66.17 & 58048.5511\\
K17T04A & 19.40 &  0.15 & 0.890 $\pm$ 0.281 & 0.739 & 0.039 (+0.061/-0.024) & 0.055 & 2.00 $\pm$ 0.50 &   0 &   1 & 85.40 & 58025.5076\\
K17U00O & 20.20 &  0.15 & 0.338 $\pm$ 0.080 & 0.290 & 0.136 (+0.089/-0.054) & 0.181 & 2.00 $\pm$ 0.50 &   1 &   1 & 81.86 & 58037.0736\\
K17U02P & 20.00 &  0.15 & 0.380 $\pm$ 0.082 & 0.347 & 0.266 (+0.136/-0.090) & 0.321 & 2.00 $\pm$ 0.50 &   6 &   6 & 73.46 & 58043.7040\\
K17U03C & 22.70 &  0.15 & 0.143 $\pm$ 0.046 & 0.117 & 0.075 (+0.057/-0.032) & 0.109 & 2.00 $\pm$ 0.50 &   1 &   1 & 87.44 & 58049.8568\\
K17V14U & 18.60 &  0.15 & 0.543 $\pm$ 0.152 & 0.485 & 0.192 (+0.230/-0.105) & 0.240 & 2.00 $\pm$ 0.50 &   2 &   2 & 76.73 & 58133.9417\\
K17V15L & 21.20 &  0.15 & 0.277 $\pm$ 0.070 & 0.252 & 0.077 (+0.044/-0.028) & 0.094 & 2.00 $\pm$ 0.50 &   6 &   6 & 74.07 & 58229.7254\\
K17W01S & 20.20 &  0.15 & 0.364 $\pm$ 0.111 & 0.319 & 0.133 (+0.094/-0.055) & 0.172 & 2.00 $\pm$ 0.50 &   7 &   7 & 79.36 & 58103.7583\\
K17W13F & 18.10 &  0.15 & 2.405 $\pm$ 0.808 & 2.326 & 0.056 (+0.044/-0.025) & 0.061 & 2.00 $\pm$ 0.50 &   7 &   7 & 61.57 & 58044.1060\\
K17W13V & 21.10 &  0.15 & 0.249 $\pm$ 0.058 & 0.239 & 0.104 (+0.111/-0.054) & 0.115 & 2.00 $\pm$ 0.50 &   0 &   3 & 63.45 & 58099.9237\\
K17W14H & 18.00 &  0.15 & 0.738 $\pm$ 0.172 & 0.682 & 0.156 (+0.081/-0.054) & 0.185 & 2.00 $\pm$ 0.50 &   5 &   5 & 71.38 & 58221.1926\\
K17W14K & 21.70 &  0.15 & 0.172 $\pm$ 0.046 & 0.165 & 0.124 (+0.139/-0.066) & 0.139 & 2.00 $\pm$ 0.50 &   0 &   4 & 65.00 & 58078.5803\\
K17X02O & 22.40 &  0.15 & 0.147 $\pm$ 0.039 & 0.136 & 0.101 (+0.060/-0.038) & 0.119 & 2.00 $\pm$ 0.50 &   2 &   2 & 71.52 & 58068.4959\\
K17Y01R & 20.00 &  0.15 & 0.351 $\pm$ 0.099 & 0.332 & 0.152 (+0.098/-0.060) & 0.173 & 2.00 $\pm$ 0.50 &   4 &   4 & 66.81 & 58120.6729\\
K17Y08S & 19.50 &  0.15 & 0.784 $\pm$ 0.209 & 0.666 & 0.052 (+0.031/-0.020) & 0.070 & 2.00 $\pm$ 0.50 &   5 &   5 & 82.99 & 58146.6197\\
K18A04E & 24.70 &  0.15 & 0.096 $\pm$ 0.017 & 0.078 & 0.025 (+0.011/-0.008) & 0.036 & 2.00 $\pm$ 0.50 &   1 &   1 & 87.51 & 58129.2398\\
K18B02Y & 20.50 &  0.15 & 0.260 $\pm$ 0.064 & 0.244 & 0.216 (+0.119/-0.077) & 0.249 & 2.00 $\pm$ 0.50 &   6 &   7 & 68.37 & 58209.4853\\
K18C02G & 19.20 &  0.15 & 0.682 $\pm$ 0.183 & 0.635 & 0.084 (+0.091/-0.044) & 0.098 & 2.00 $\pm$ 0.50 &   3 &   3 & 70.08 & 58161.2419\\
K18C02O & 19.40 &  0.15 & 0.516 $\pm$ 0.141 & 0.451 & 0.099 (+0.063/-0.038) & 0.128 & 2.00 $\pm$ 0.50 &   2 &   2 & 79.48 & 58184.8028\\
K18C02Z & 21.40 &  0.15 & 0.272 $\pm$ 0.062 & 0.225 & 0.071 (+0.053/-0.030) & 0.100 & 2.00 $\pm$ 0.50 &   1 &   1 & 85.88 & 58181.3388\\
K18C14B & 21.30 &  0.15 & 0.223 $\pm$ 0.054 & 0.213 & 0.108 (+0.064/-0.040) & 0.120 & 2.00 $\pm$ 0.50 &   0 &   4 & 63.83 & 58159.6126\\
K18E00E & 21.70 &  0.15 & 0.188 $\pm$ 0.043 & 0.178 & 0.131 (+0.071/-0.046) & 0.149 & 2.00 $\pm$ 0.50 &   5 &   5 & 67.12 & 58187.7701\\
K18J01E & 19.90 &  0.15 & 0.375 $\pm$ 0.111 & 0.368 & 0.168 (+0.125/-0.072) & 0.178 & 2.00 $\pm$ 0.50 &   5 &   5 & 56.72 & 58280.5585\\
K18L05F & 20.20 &  0.15 & 0.310 $\pm$ 0.102 & 0.299 & 0.139 (+0.107/-0.060) & 0.153 & 2.00 $\pm$ 0.50 &  20 &  19 & 62.55 & 58309.0224\\
K18L15Q & 20.80 &  0.15 & 0.377 $\pm$ 0.082 & 0.334 & 0.166 (+0.081/-0.054) & 0.210 & 2.00 $\pm$ 0.50 &  12 &  13 & 77.66 & 58309.5869\\
K18M06Y & 21.20 &  0.15 & 0.669 $\pm$ 0.207 & 0.612 & 0.038 (+0.027/-0.016) & 0.046 & 2.00 $\pm$ 0.50 &  11 &  12 & 72.91 & 58293.4605\\
K18M06Y & 21.20 &  0.15 & 0.759 $\pm$ 0.182 & 0.676 & 0.040 (+0.022/-0.014) & 0.050 & 2.00 $\pm$ 0.50 &   8 &   8 & 76.86 & 58281.9182\\
K18T04D & 22.10 &  0.15 & 0.250 $\pm$ 0.061 & 0.208 & 0.048 (+0.026/-0.017) & 0.067 & 2.00 $\pm$ 0.50 &   3 &   3 & 85.56 & 58561.3594\\
K18U01A & 19.70 &  0.15 & 0.374 $\pm$ 0.087 & 0.336 & 0.162 (+0.084/-0.055) & 0.201 & 2.00 $\pm$ 0.50 &   3 &   3 & 75.72 & 58407.8800\\
K18U01N & 19.30 &  0.15 & 0.649 $\pm$ 0.178 & 0.642 & 0.096 (+0.076/-0.042) & 0.101 & 2.00 $\pm$ 0.50 &   4 &   4 & 53.61 & 58446.8304\\
K18U02Q & 21.90 &  0.15 & 0.273 $\pm$ 0.076 & 0.234 & 0.055 (+0.035/-0.021) & 0.074 & 2.00 $\pm$ 0.50 &   2 &   2 & 81.81 & 58408.2443\\
K18V00X & 21.10 &  0.15 & 0.168 $\pm$ 0.035 & 0.144 & 0.212 (+0.174/-0.096) & 0.281 & 2.00 $\pm$ 0.50 &   1 &   1 & 81.40 & 58419.1726\\
K18W01R & 20.30 &  0.15 & 0.458 $\pm$ 0.160 & 0.445 & 0.084 (+0.069/-0.038) & 0.091 & 2.00 $\pm$ 0.50 &   3 &   3 & 60.52 & 58484.7210\\
K18X00R & 18.50 &  0.15 & 0.710 $\pm$ 0.173 & 0.645 & 0.137 (+0.081/-0.051) & 0.166 & 2.00 $\pm$ 0.50 &   1 &   1 & 74.03 & 58500.9948\\
K18X02Z & 21.60 &  0.15 & 0.259 $\pm$ 0.071 & 0.234 & 0.196 (+0.123/-0.075) & 0.241 & 2.00 $\pm$ 0.50 &  34 &  34 & 75.09 & 58493.9276\\
K19C02D & 20.09 &  0.15 & 0.441 $\pm$ 0.127 & 0.377 & 0.082 (+0.067/-0.037) & 0.110 & 2.00 $\pm$ 0.50 &   1 &   1 & 82.26 & 58518.3875\\
K19C02L & 23.34 &  0.15 & 0.073 $\pm$ 0.018 & 0.063 & 0.179 (+0.099/-0.064) & 0.234 & 2.00 $\pm$ 0.50 &   1 &   1 & 80.40 & 58558.8947\\
K19E00Q & 19.36 &  0.15 & 0.473 $\pm$ 0.131 & 0.459 & 0.128 (+0.092/-0.054) & 0.139 & 2.00 $\pm$ 0.50 &   7 &   7 & 60.65 & 58544.3332\\
K19F00D & 23.32 &  0.15 & 0.215 $\pm$ 0.062 & 0.173 & 0.019 (+0.024/-0.011) & 0.028 & 2.00 $\pm$ 0.50 &   1 &   1 & 88.69 & 58558.9033\\
K19J05U & 24.00 &  0.15 & 0.056 $\pm$ 0.011 & 0.048 & 0.128 (+0.103/-0.057) & 0.173 & 2.00 $\pm$ 0.50 &   3 &   3 & 82.75 & 58607.7321\\
K19N07D & 21.43 &  0.15 & 0.198 $\pm$ 0.048 & 0.165 & 0.122 (+0.100/-0.055) & 0.170 & 2.00 $\pm$ 0.50 &   0 &   2 & 84.82 & 58677.9404\\

\label{tab.props}
\end{longtable}
 }


\begin{thebibliography}{XXX}

\bibitem[Astropy Collaboration \etal(2013)]{astropy1}
  Astropy Collaboration, Robitaille, T.~P., Tollerud, E.~J., \etal, 2013, A\&A, 558, A33.

\bibitem[Astropy Collaboration \etal(2018)]{astropy2}
  Astropy Collaboration, Price-Whelan, A.~M., Sip{\H{o}}cz, B.~M., \etal, 2018, AJ, 156, 123.
  
 \bibitem[Cutri \etal(2015)]{cutri15}
 Cutri, R.M., Mainzer, A., Conrow, T., Masci, F., Bauer, J., \etal, 2015, Explanatory Supplement to the NEOWISE Data Release Products, {\it https://wise2.ipac.caltech.edu/docs/release/neowise/expsup}

\bibitem[Delbo' \etal(2007)]{delbo07}
  Delbo', M., dell'Oro, A., Harris, A.W., Mottola, S., \& Mueller, M., 2007, Icarus, 190, 236.

\bibitem[Delbo \etal(2015)]{delbo15}
Delbo', M., Mueller, M., Emery, J.P., Rozitis, B., \& Capria, M.T., 2015, Asteroids IV (P. Michel, F. DeMeo, W.F. Bottke eds), University of Arizona Press, 107.

\bibitem[Hanu\u{s} \etal(2016)]{hanus16}
Hanu\u{s}, J., Delbo', M., Vokrouhlick\'{y}, D., 2016, A\&A, 592, 34.
  
\bibitem[Harris \etal(1998)]{harris98}
Harris, A.W., 1998, Icarus, 131, 291.

\bibitem[Koren \etal(2015)]{koren15}
  Koren, S.C., Wright, E.L. \& Mainzer, A.K., 2015, Icarus, 258, 82.

\bibitem[Mainzer \etal(2011a)]{mainzer11}
Mainzer, A.K., Bauer, J.M., Grav, T., Masiero, J., \etal, 2011a, ApJ, 731, 53.
 
\bibitem[Mainzer \etal(2011b)]{mainzer11neo}
Mainzer, A.K., Grav, T., Bauer, J.M., Masiero, J., \etal, 2011b, ApJ, 743, 156.
 
\bibitem[Mainzer \etal(2014a)]{mainzer14neowise}
Mainzer, A.K., Bauer, J., Cutri, R., Grav, T., Masiero, J., \etal, 2014a, ApJ, 792, 30.
 
\bibitem[Mainzer \etal(2014b)]{mainzer14tinyneo}
Mainzer, A.K., Bauer, J., Grav, T., Masiero, J., Cutri, R.,  \etal, 2014b, ApJ, 784, 110.
 
\bibitem[Masiero \etal(2017)]{masiero17}
Masiero, J.R, Nugent, C., Mainzer, A.K., Wright, E., Bauer, J., \etal, 2017, AJ, 154, 168.

\bibitem[Masiero \etal(2018)]{masiero18}
Masiero, J.R, Redwing, E., Mainzer, A.K., Bauer, J.M., Cutri, R.M., \etal, 2018, AJ, 156, 60.

\bibitem[Masiero \etal(2019)]{masiero19}
Masiero, J.R, Wright, E.L., \& Mainzer, A.K., 2019, AJ, 158, 97.

\bibitem[Masiero \etal(2020)]{masieroY45}
Masiero, J.R, Mainzer, A.K., Grav, T., \etal, 2020, PSJ, 1, 5.
 
\bibitem[Mommert \etal(2018)]{mommert18}
Mommert, M., Jedicke, R., \& Trilling, D., 2018, AJ, 155, 74.

\bibitem[Nugent \etal(2015)]{nugent15}
Nugent, C.R., Mainzer, A., Masiero, J., Bauer, J.M., Cutri, R.M. \etal, 2015, ApJ, 814, 117.

\bibitem[Nugent \etal(2016)]{nugent16}
  Nugent, C.R., Mainzer, A., Bauer, J.M., Cutri, R.M., Kramer, E. \etal, 2016, AJ, 152, 63.

\bibitem[Trilling \etal(2016)]{trilling16}
  Trilling, D.E., Mommert, M., Hora, J., \etal, 2016, AJ, 152, 172.

\bibitem[Virtanen \etal(2020)]{scipy}
  Virtanen, P., Gommers, R., Oliphant, T., \etal, 2020, Nature Methods, 17, 261.
  
\bibitem[Wright \etal(2010)]{wright10}
Wright, E.L., Eisenhardt, P., Mainzer, A.K., Ressler, M.E., Cutri, R.M., \etal, 2010, AJ, 140, 1868.

\bibitem[Wright \etal(2018)]{wright18}
Wright, E.L., Mainzer, A.K., Masiero, J., Grav, T., Cutri, R.M., Bauer, J.M., 2018, arXiv:1811.01454.


\clearpage

\end{thebibliography}
\end{document}